\documentclass[letterpaper,twocolumn,10pt]{article}
\usepackage{usenix-2020-09}
\usepackage{pgfplots}
\pgfplotsset{compat=1.8}
\usepgfplotslibrary{fillbetween}
\usepackage{tikz}
\usetikzlibrary{matrix,positioning}
\usetikzlibrary{external}
\usepackage{pgfplots}
\usepackage{pgfplotstable}
\usetikzlibrary{pgfplots.groupplots}
\usetikzlibrary{arrows}
\usetikzlibrary{patterns}
\usetikzlibrary{positioning}
\usetikzlibrary{decorations.pathreplacing}
\usetikzlibrary{shapes.arrows}
\usetikzlibrary{shapes.geometric,shapes.misc}
\usetikzlibrary{pgfplots.groupplots}
\pgfplotsset{compat=newest}
\pgfkeys{/pgf/number format/.cd,1000 sep={}}

\pgfplotsset{
    discard if/.style 2 args={
        x filter/.code={
            \edef\tempa{\thisrow{#1}}
            \edef\tempb{#2}
            \ifx\tempa\tempb
                
            \fi
        }
    },
    discard if not/.style 2 args={
        x filter/.code={
            \edef\tempa{\thisrow{#1}}
            \edef\tempb{#2}
            \ifx\tempa\tempb
            \else
                
            \fi
        }
    }
}

\makeatletter% from https://tex.stackexchange.com/a/13665
\newcommand\resetstackedplots{%
\pgfplots@stacked@isfirstplottrue
}
\makeatother

\usetikzlibrary{spy}
\usetikzlibrary{patterns,backgrounds}
\pgfdeclarelayer{foreground}
\pgfsetlayers{background,main,foreground}

\usepgfplotslibrary{statistics}

\definecolor{kaito}{RGB}{0, 128, 255}

\usepackage[normalem]{ulem}
\usepackage{enumitem}
\usepackage{rotating}
\usepackage{hyperref}
\usepackage{booktabs}
\usepackage{balance}
\usepackage{graphicx}
\usepackage{wrapfig}
\usepackage[ruled, vlined, linesnumbered]{algorithm2e}
\usepackage{amsmath}
\usepackage{fancyhdr}
\usepackage{bbm}
\usepackage{multirow}
\usepackage{xcolor}
\usepackage{soul}
\usepackage{tikz}
\usepackage{cleveref}
\usepackage{url}

\usepackage{mathtools}
\usepackage{pifont}
\usepackage{makecell}
\usepackage{subcaption}
\usepackage{xspace}
\usepackage[skins]{tcolorbox}
\usepackage{array}
\usepackage[super]{nth}

\newcommand{\PreserveBackslash}[1]{\let\temp=\\#1\let\\=\temp}
\newcolumntype{C}[1]{>{\PreserveBackslash\centering}p{#1}}
\newcolumntype{R}[1]{>{\PreserveBackslash\raggedleft}p{#1}}
\newcolumntype{L}[1]{>{\PreserveBackslash\raggedright}p{#1}}
\def\BibTeX{{\rm B\kern-.05em{\sc i\kern-.025em b}\kern-.08em
    T\kern-.1667em\lower.7ex\hbox{E}\kern-.125emX}}

\setlist{noitemsep, leftmargin=*, topsep=0pt, partopsep=0pt}

\crefformat{section}{\S#2#1#3} % see manual of cleveref, section 8.2.1
\crefformat{subsection}{\S#2#1#3}
\crefformat{subsubsection}{\S#2#1#3}

\tcbset{
  takeaway/.style={ enhanced,width=\hsize,left=0pt,right=0pt,top=0pt,bottom=0pt,colback=black!20!blue!4!white,boxrule=1pt,colframe=black!30!blue!20!white
  },
}
\newcounter{takeawaycounter}
\newcommand{\takeawaybox}[1]{
    \stepcounter{takeawaycounter}
    \begin{tcolorbox}[takeaway]
    \textbf{Takeaway~\arabic{takeawaycounter}:}#1
    \end{tcolorbox}
}

\tcbset{
  conclusion/.style={ enhanced,width=\hsize,left=0pt,right=0pt,top=0pt,bottom=0pt,colback=black!20!green!4!white,boxrule=1pt,colframe=black!30!green!20!white
  },
}
\newcounter{conclusioncounter}
\newcommand{\conclusionbox}[1]{
    \stepcounter{conclusioncounter}
    \begin{tcolorbox}[conclusion]
    \textbf{Carbon Implication~\arabic{conclusioncounter}:}#1
    \end{tcolorbox}
}

\newcommand{\SYSTEM}{GreenLLM}
\newcommand{\DistPD}{Disg-Pref-Decode}
\newcommand{\DistSD}{Disg-Spec-Decode}

\usepackage{authblk}

\begin{document}

% \title{\SYSTEM{}: Reducing Carbon Emissions in Large Language Model Serving \\ by Reusing Older GPUs}

\title{\SYSTEM{}: Disaggregating Large Language Model Serving on \\Heterogeneous GPUs for Lower Carbon Emissions}

\author[1,*]{Tianyao Shi}
\author[1,*]{Yanran Wu}
\author[2]{Sihang Liu}
\author[1]{Yi Ding}
\affil[1]{Purdue University}
\affil[2]{University of Waterloo}

% Disaggregating Large Language Model Serving \\ on Heterogeneous GPUs with Lower Carbon

% \SYSTEM{}: Achieve Low Carbon Emissions with Disaggregating Large Language Model Serving using Heterogeneous GPUs

%Designing Sustainable Large Language Model Serving \\ with Lower Carbon

\maketitle
\def\thefootnote{*}\footnotetext{These authors contributed equally.}\def\thefootnote{\arabic{footnote}}

\begin{abstract}	

LLMs have been widely adopted across many real-world applications. However, their widespread use comes with significant environmental costs due to their high computational intensity and resource demands. Specifically, this has driven the development of new generations of high-performing GPUs, exacerbating the problem of electronic waste and accelerating the premature disposal of devices.  

To address this problem, this paper focuses on reducing the carbon emissions of LLM serving by reusing older, low-performing GPUs. We present \SYSTEM{}, an SLO-aware LLM serving framework designed to minimize carbon emissions by reusing older GPUs. \SYSTEM{} builds on two identified use cases that disaggregate specific computations onto older GPUs, reducing carbon emissions while meeting performance goals. To deepen our understanding of the potential carbon savings from disaggregation, we also provide a theoretical analysis of its relationship with carbon intensity and GPU lifetime. Our evaluations show that \SYSTEM{} reduces carbon emissions by up to 40.6\% compared to running standard LLM serving on new GPU only, meeting latency SLOs for over 90\% of requests across various applications, latency requirements, carbon intensities, and GPU lifetimes.

% Evaluation on various applications, latency requirements, carbon intensities, and GPU lifetime show that \SYSTEM{} can save up to 40.6\% carbon emissions, compared to the standard configuration, while satisfying the SLO constraints for over 90\% of requests.

\end{abstract}

\section{Introduction}\label{sec:intro}

Large language models (LLMs) have been widely adopted in various real-world applications for their ability to perform complex natural language processing tasks~\cite{vu2023freshllms,shen2024hugginggpt,liu2024your}. Despite their widespread use, serving LLMs causes significant environmental impact in terms of carbon emissions~\footnote{Detailed definitions of various carbon emissions are provided in~\cref{subsec:carbon-account}.}. For example, processing a single ChatGPT prompt generates over 4 grams of CO\textsubscript{2}eq~\cite{chatgptcarbon2023}, which is more than 20 times the \emph{operational carbon emission} of a web search query~\cite{whyyourinternet2020}. 
% Futhermore, serving LLMs relies on high-performing but power-hungry GPUs, which leads to high embodied carbon emissions. 
% is highly energy-intensive, leading to significant carbon emissions due to the energy usage to power the datacenter. This type of carbon emission is referred to as the \emph{operational carbon emissions}. 

% On the other hand, recent research on sustainable artificial intelligence (AI) highlights that \emph{embodied carbon emissions} from the device manufacturing process dominate the lifecycle carbon emissions of computing systems~\cite{wu2022sustainable,gupta2021chasing}. This issue is especially pronounced in generative AI like LLM serving~\cite{chien2023reducing}, as building the infrastructure for LLMs highly depends on high-end GPUs like A100 and H100. The demand for these GPUs drives a surge in manufacturing and production, leading to substantial embodied carbon emissions, particularly for devices with short life cycles~\cite{gupta2021chasing}. 

Serving LLMs relies on high-performing but power-hungry GPUs. The rising demand for computation has driven the development of new generations of high-performing GPUs, whose manufacturing processes contribute significantly to environmental impact in terms of \emph{embodied carbon emissions}~\cite{wu2022sustainable,gupta2021chasing}. Meanwhile, older GPUs become increasingly overlooked and inadequate for running the latest LLM serving systems, further exacerbating the growing problem of electronic waste (e-waste)~\cite{ai_e-waste_2024,ieee_spectrum_ewaste2024}. 
Recent studies estimate that e-waste from generative artificial intelligence (AI) could reach 1.2 to 5.0 million tons between 2020 and 2030~\cite{wang2024waste}, with rapid advances in AI that accelerate obsolescence and premature disposal of devices~\cite{ewaste2024,switzer2023junkyard}. 

In this paper, we tackle the challenge of reducing the total carbon emissions of LLM serving by reusing old, low-performing GPUs.  
Since embodied carbon emissions are amortized over a device's lifetime, extending hardware lifetimes by reusing older GPUs will effectively reduce these emissions~\cite{gupta2022act}. To this end, we explore GPU heterogeneity to enable the reuse of older GPUs for serving LLMs, while meeting performance goals. Specifically, we investigate strategies for disaggregating a portion of the computation of LLM serving to older GPUs to reduce total carbon emissions.

We first explore allocating the prefill and decoding phases of LLM serving onto different types of GPUs. LLM serving operates in two phases---prefill and decoding (as detailed in~\cref{subsec:llm}). The prefill phase processes the prompt, stores the context in the key-value (KV) cache, and is compute-bound, while the decoding phase generates tokens autoregressively and is memory-bound. Prior work has either separated the prefill and decoding phases on homogeneous GPUs to eliminate interference and improve performance~\cite{zhong2024distserve,hu2024TetriInfer}, or heterogeneous GPUs to reduce power and costs~\cite{patel2024splitwise,griggs2024melange}. In contrast, we minimize carbon emissions by using heterogeneous GPUs, especially old, low-performing GPUs.
We allocate the prefill phase to newer, more advanced GPU for faster prompt processing, while assigning the decoding phase to older, less advanced GPU to reduce embodied carbon emissions. We call this approach \textbf{\DistPD{}}, and our experiments show that it can effectively reduce total carbon emissions while meeting latency service level objectives (SLOs).

While \DistPD{} demonstrates a feasible use case to reduce carbon emissions for LLM serving, it has a key limitation: the high network bandwidth requirements due to the large KV cache transfer between GPUs. 
% For instance, disaggregating the prefill and decoding phases of a Llama-7B model on A100 and T4 GPUs with 16Gbps bandwidth can quickly exceed memory limits at request rates below 2 (\Cref{fig:bandwidth}). 
For instance, disaggregating the prefill and decoding phases of a Llama-7B model on A100 and T4 GPUs requires bandwidth over 10Gbps even when the request rate of 1 prompt/s (see \Cref{fig:bandwidth}). 
In practice, the carbon reduction benefits of \DistPD{} can diminish as the KV cache size increases, particularly because the network devices connecting heterogeneous GPUs can have slower data transfer speeds. This overhead can offset the gains achieved by \DistPD{} when high network bandwidth is not available.

To address this limitation, we explore a second disaggregation use case based on speculative decoding for systems without high-bandwidth interconnect, while still reducing total carbon emissions. Speculative decoding accelerates LLM serving by running two LLM models concurrently~\cite{leviathan2023fast,miao2024specinfer}: a larger, slower model and a smaller, faster model. Instead of generating tokens sequentially as in standard LLMs, the smaller model generates a set of tokens, and the larger model verifies them in parallel. This approach reduces the computational load on the larger model by offloading initial predictions to the smaller model, allowing it to focus primarily on validation. 
Building on this concept, we introduce a disaggregated version of speculative decoding (\textbf{\DistSD{}}) using heterogeneous GPUs. In \DistSD{}, the larger model is assigned to the newer GPU, while the smaller model runs on the older GPU. Unlike \DistPD{}, \DistSD{} has a low network bandwidth requirement since it only transfers speculative tokens and the corresponding probability distribution between GPUs.
% as validated by our experiments.  

Building on the two use cases we identified, we design and implement \SYSTEM{}, an SLO-aware LLM serving system that minimizes total carbon emissions by leveraging heterogeneous GPUs. \SYSTEM{} includes three components: a disaggregated system supporting various LLM serving configurations such as \DistPD{} and \DistSD{}; a profiler measuring performance and energy consumption across various request sizes and rates; and an SLO-aware scheduler that searches for the optimal configuration for minimal carbon emissions while meeting latency SLOs. 

To further understand disaggregation's carbon savings, we theoretically compare standalone (new GPU only) and disaggregated (new and old GPUs) setups. We address three questions: 1) When does disaggregation yield greater carbon savings? 2) How does carbon intensity influence carbon savings? 3) How does GPU lifetime influence carbon savings?
 
We evaluate \SYSTEM{} on three real-world LLM applications: chatbots, programming assistants, and document summarization. Compared to running standard LLM serving on new GPU only, \SYSTEM{} reduces carbon emissions by up to 40.6\% while meeting latency SLOs. Additionally, \SYSTEM{} achieves up to 27.9\% carbon savings across regions and GPU lifetimes, even in low-carbon intensity settings (17 gCO$_2$/kWh). 
We summarize the contributions of this paper as follows:

\begin{itemize}
\item We leverage GPU heterogeneity to explore the reusing of older GPUs for LLM serving to reduce carbon emissions.
\item We identify two use cases, \DistPD{} and \DistSD{}, to enable the disaggregation of LLM serving on heterogeneous GPUs.  
\item We address high bandwidth constraints by introducing a disaggregated speculative decoding approach with low network bandwidth requirements.  
\item We theoretically analyze the carbon emissions of disaggregated serving, highlighting its relationship with carbon intensity and GPU lifetime.
\item We design and implement \SYSTEM{}, an SLO-aware LLM serving system that minimizes carbon emissions by reusing old GPUs. We evaluate \SYSTEM{} using three LLM serving datasets, validating our theoretical analysis. 
\end{itemize}

\section{Background and Related Work}\label{sec:background}

\begin{figure}[!t]
	\begin{center}
		\includegraphics[width=\linewidth]{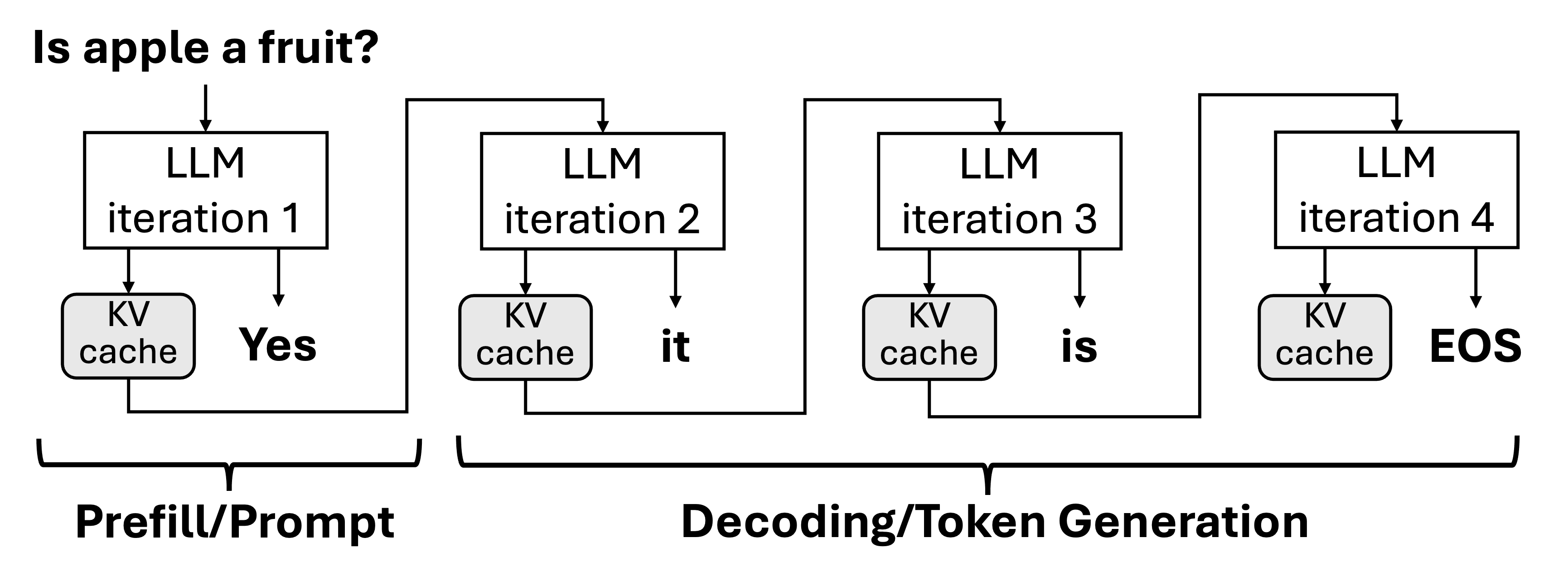}  \vspace{-0.1in}
	\end{center}\vspace{-0.3in}
	\caption{\textbf{An LLM serving example.}}\label{fig:llm}
\end{figure}

This section first introduces the LLM inference process. We then review recent work on disaggregated serving and speculative decoding as potential use cases. Then, we discuss carbon accounting methods for operational and embodied emissions, followed by a literature survey of recent carbon reduction efforts in various applications and systems.

\subsection{LLM Serving Process}\label{subsec:llm}

LLMs are deep learning generative models based on transformer architecture~\cite{vaswani2017attention}. Transformers use attention and multi-layer perceptron layers to process inputs and generate outputs. Figure~\ref{fig:llm} shows an example of LLM serving. Upon receiving a prompt request, all input tokens are processed in parallel within a single iteration to generate the first token. We call this the \emph{prefill} or prompt processing phase. The context generated by the attention layers during prefill is stored in the key-value (KV) cache~\cite{shoeybi2019megatron,yu2022orca}, as it is required for all subsequent token generation iterations. 
After the first token is generated, the following tokens are generated using only the last generated token and the KV-cache as inputs for the model's forward pass. We call this the \emph{decoding} or token generation phase, which demands more memory bandwidth and capacity compared to the computationally intensive prefill phase. Furthermore, each phase has its own specific latency SLOs and preferences for different types of parallelism~\cite{zhong2024distserve}.

\paragraph{Performance metrics.} LLM serving performance is mainly evaluated using two latency metrics: TTFT (time-to-first-token) and TPOT (time-per-output-token)~\cite{zhong2024distserve,patel2024splitwise,nvidia-llm-metrics,databricks-llm-inference}. 
TTFT quantifies the time it takes to generate the first output token after a request enters the system, reflecting the model's responsiveness during the prefill phase. TPOT measures the time interval between generating consecutive output tokens, which represents performance during the decoding phase.

\subsection{LLM Serving Performance Optimization}\label{subsec:related}

Various techniques and open-source platforms, such as vLLM~\cite{kwon2023efficient}, have been developed to accelerate LLM serving. Some techniques, like request batching~\cite{yu2022orca,agrawal2024taming} and parallelism~\cite{brakel2024model}, have become standard. Here, we review the following two techniques that we identify as use cases.

\paragraph{Disaggregated prefill and decoding.} This method improves LLM serving performance by assigning the prefill and decoding phases to separate GPUs, rather than co-locating them~\cite{zhong2024distserve,patel2024splitwise,hu2024TetriInfer,griggs2024melange}. This approach offers two benefits: (1) eliminating interference between the two phases, and (2) enabling tailored resource allocation and parallelism strategies for each phase. However, disaggregation requires high network bandwidth to transfer the KV cache between GPUs. While modern GPU clusters with high-bandwidth networking like NVLink or Infiniband can mitigate this overhead, such environments may not always be practical. Furthermore, while prior works use multiple GPUs, \emph{they solely rely on high-end GPUs and neglect the potential of using older GPUs to reduce embodied carbon emissions.} 

% \begin{figure}[!t]
% 	\begin{center}
% 		\includegraphics[width=\linewidth]{fig/spec}  \vspace{-0.1in}
% 	\end{center}\vspace{-0.3in}
% 	\caption{Illustration of speculative decoding.}\label{fig:spec}
% \end{figure}

\paragraph{Speculative decoding.} This method consists of three primary components: the \textit{draft model}, the \textit{target model}, and the \textit{verifier}~\cite{leviathan2023fast,miao2024specinfer}. The draft model is a smaller, more efficient model that generates a sequence of speculative tokens, while the target model, which is larger and more capable, takes these speculative tokens and performs decoding to compute the probability distribution. Finally, the verifier decides whether to accept the tokens proposed by the draft model by comparing the probabilities from both models, typically using a rejection sampling scheme. This draft-then-verify paradigm enables simultaneous decoding of multiple tokens per step, significantly accelerating inference compared to the traditional autoregressive decoding.

The verifier employs a rejection sampling mechanism to determine whether to accept speculative tokens. Specifically, given a sequence of tokens $\{x_1, \dots, x_n\}$ and a set of $K$ draft tokens $\{\tilde{x}_{n+1}, \dots, \tilde{x}_{n+K}\}$ generated from the draft model's conditional distribution $p(\cdot | x_1, \dots, x_n)$, the token $\tilde{x}_{n+1}$ is accepted with probability:
\[
\min \left( 1, \frac{q(\tilde{x}_{n+1} | x_1, \dots, x_n)}{p(\tilde{x}_{n+1} | x_1, \dots, x_n)} \right),
\]
where $q(\tilde{x}_{n+1} | x_1, \dots, x_n)$ and $p(\tilde{x}_{n+1} | x_1, \dots, x_n)$ are the probabilities of the token $\tilde{x}_{n+1}$ from target and draft models, respectively, conditioned on the prior tokens in the sequence. Thus, there is a temporal dependency between these components: the target model relies on the speculative tokens predicted by the draft model, and the verifier depends on the probability outputs from both models. Prior work co-locates both models on the same GPU, \emph{neglecting the potential of assigning the draft model to an older, less resource-intensive GPU to further optimize for carbon emissions.}

% This method accelerates LLM serving by running two LLM models concurrently~\cite{leviathan2023fast,miao2024specinfer}: a draft (smaller, faster) model and a target (larger, slower) model. Instead of generating tokens sequentially as in standard LLMs, the draft model first generates a set of tokens rapidly. The target model then processes these tokens and performs decoding to compute their probability distribution. Finally, a verifier decides whether to accept the tokens proposed by the draft model by comparing the probabilities from both models, typically using a rejection sampling scheme~\cite{chen2023accelerating}. This draft-then-verify approach enables simultaneous decoding of multiple tokens per step, significantly accelerating inference compared to traditional autoregressive decoding. However, the prior work co-locates both models on the same GPU, \emph{neglecting the potential of assigning the draft model to an older, less resource-intensive GPU to further optimize for carbon emissions.}

\paragraph{Summary.} While various approaches have been introduced to optimize LLM service, they focus on improving performance and using high-end GPUs. This paper aims to explore how disaggregation techniques on heterogeneous GPUs can be harnessed to reduce total carbon emissions.

\subsection{Carbon Emissions Accounting}\label{subsec:carbon-account}

According to the Greenhouse Gas Protocol~\cite{protocol2011greenhouse}, carbon emissions are classified into three categories: Scope 1 (direct emissions), Scope 2 (indirect emissions from energy consumption), and Scope 3 (indirect emissions from the production and transportation of goods procured). Following the terminology used in prior work~\cite{gupta2021chasing,gupta2022act}, we refer to Scope 1 as direct carbon emissions, Scope 2 as operational carbon emissions, and Scope 3 as embodied carbon emissions. In this paper, the total carbon emissions that we model and evaluate is the sum of embodied and operational carbon emissions.

\paragraph{Embodied carbon.} The embodied carbon of a hardware device $C_{\rm e}$ is modeled based on the processor chip areas and memory capacities~\cite{gupta2022act,faiz2024llmcarbon}. To illustrate, we compare the specifications and embodied carbon emissions of three GPU types---T4, V100, and A100---in~\Cref{tbl:spec}. For the rest of this paper, we will conduct experiments using these three GPU types. The A100, the latest-generation GPU, has the greatest compute capability, semiconductor technology, the largest chip area, and memory capacity, leading to the highest embodied carbon. T4 and V100 were introduced earlier and have the same memory capacity. Because V100 has a larger chip area, it has higher embodied carbon than T4. As such, the choice of GPU type can significantly influence the embodied carbon emission of a computing task.

\begin{table}[!t]
	\footnotesize
	\begin{center}
		\caption{\textbf{Specifications of GPUs in this paper.}}
		\begin{tabular}{llll}\toprule
			GPU Model                  & T4          & V100           & A100        \\ \midrule
			VRAM (GB)                  & 16          & 16             & 40             \\
			Memory Bandwidth (GB/s)    & 320         & 900            & 1555           \\
			Chip Area (mm$^2$) & 545   & 815   & 826       \\
			Max Power (W)              & 70          & 300            & 400            \\
			Technology Node (nm)              & 12    & 12    & 7         \\
			FP16 TFLOPs                & 65          & 28.26          & 312            \\
			%		TF32 TFLOPs                & 8.141       & 15.7           & 156            \\
			Year                       & 2018        & 2017           & 2020           \\ \hline
			%		Price (\$/GPU hour on GCP) & 0.35        & 2.48           & 3.67           \\
			\textbf{Embodied Carbon (kgCO\textsubscript{2})}       & \textbf{10.3}        & \textbf{20}             & \textbf{26.34}        
			\\ \bottomrule    
		\end{tabular}\label{tbl:spec}
	\end{center}
\end{table}

The embodied carbon emission of a request is amortized over time by the ratio of the total execution time to the hardware lifetime ($\texttt{LT}$)~\cite{gupta2022act}. The lifetime of a GPU is 5-7 years~\cite{ostrouchov2020gpulife}. Therefore, a request executed on a GPU for $t_{\rm req}$ time generates the embodied carbon emission of:
\begin{align}\label{eq:carbon-eb}
	C_{\rm req,e} = \frac{t_{\rm req}}{\texttt{LT}}\cdot{C_{\rm e}}
\end{align}
\paragraph{Operational carbon.} The operational carbon emission of a request $C_{\rm req,o}$ is determined by the product of its energy consumption $E_{\rm request}$ and carbon intensity of the grid ($\texttt{CI}$). \emph{Carbon intensity} is measured as grams of $CO_{\rm 2eq}$ emitted per $kWh$ of electricity generated or consumed~\cite{maji2022carboncast,li2024uncertainty}. Therefore, the operational carbon emission of a request is:
\begin{align}\label{eq:carbon-op}
	C_{\rm req,o} = E_{\rm req}\cdot \texttt{CI}
\end{align}
\noindent \textbf{Total carbon.} The total carbon emission consists of embodied and operational carbon emissions. For a request that executes for $t_{\rm request}$ time, the total carbon emission is:
\begin{align}\label{eq:carbon-total}
	C_{\rm req} = C_{\rm req,e} + C_{\rm req,o} 
	= \frac{t_{\rm req}}{\texttt{LT}}{C_{\rm e}}+E_{\rm req}\cdot \texttt{CI}
\end{align}

\subsection{Carbon Emission Reduction}\label{subsec:carbon-opt}

There are two main lines of research aimed at reducing carbon emissions: one focuses on reducing operational carbon and the other on reducing embodied carbon.  

Reducing operational carbon focuses on load shifting based on carbon intensity across regions and times~\cite{radovanovic2022carbon}. Carbon Explorer provides a holistic framework for reducing carbon emissions for datacenters~\cite{acun2023carbon}. Ecovisor introduces a carbon-efficient virtual energy management system~\cite{souza2023ecovisor}. CarbonScaler uses workload elasticity to reduce operational carbon~\cite{hanafy2024carbonscaler}. GAIA optimizes the cost of reducing cloud carbon emissions~\cite{hanafy2024going}. Caribou reduces carbon emissions from serverless applications through geospatial shifting~\cite{gsteiger2024caribou}.  

Reducing embodied carbon focuses on extending hardware lifetimes. Junkyard Computing repurposes discarded smartphones to reduce carbon footprints for conventional cloud applications with low compute intensity, such as microservices~\cite{switzer2023junkyard}. GreenSKU reconfigures old, functional non-GPU hardware devices in datacenters, such as DRAMs and SSDs, for new servers~\cite{wang2024designing}. However, these approaches target reusing non-GPU hardware for low-compute-intensive workloads and are not directly applicable to LLM serving, which demands significantly higher compute capabilities.  

Recently, some work has begun exploring sustainable LLM serving through carbon characterization and analysis~\cite{nguyen2024towards,li2024towards,chien2023reducing,faiz2024llmcarbon}.
Despite these efforts, to the best of our knowledge, there are no system-level solutions for carbon reduction in LLM serving, nor do they leverage disaggregation techniques to reuse older GPUs~\cite{li2024genai}.

%\noindent \textbf{(1) Batching.} Batching techniques group multiple requests to improve the throughput of LLM serving. Naive batching operates at the request level, where ready requests are batched and processed together, completing all batched requests before starting new ones. ORCA improves upon this by introducing continuous batching, which selectively batches specific operations, such as requests in the prefill phase or the decoding phase~\cite{yu2022orca}. Sarathi-Serve introduces chunked prefills, dividing large prefill operations into smaller chunks that are processed across multiple iterations~\cite{agrawal2023sarathi}.  
%
%\noindent \textbf{(2) Parallelism.} Model parallelism, specifically pipeline (PP) and tensor (TP) parallelism, can distribute LLM models across multiple GPUs or machines to improve efficiency and memory capacity~\cite{brakel2024model}. PP partitions layers across GPUs, minimizing communication, while TP divides tensors across GPUs, requiring high-bandwidth interconnects.

\section{Motivation}\label{sec:motivation}

Serving LLMs on new GPUs incurs significant embodied and total carbon emissions. To reduce embodied carbon, we explore the potential of offloading a portion of the computation to older GPUs. 
However, this approach faces two technical challenges. First, offloading parts of the LLM model, such as the decoding phase, to older GPUs may lead to performance degradation. Second, the communication overhead between new and old GPUs, particularly when transferring KV cache, can be substantial. In this section, we characterize these challenges to identify potential solutions.

For the following experiments, we use randomized texts that match the input and output token lengths, as our performance and carbon emissions evaluations are independent of the text but only concerned with token lengths. 
% as the model output quality does not influence the problem we are investigating.
We test on three GPU types: A100, V100, and T4.

\subsection{Model Size and Phase Matter}\label{subsec:model-size}

We characterize latency and energy consumption for both the prefill and decoding phases of the Llama model in various sizes (300M, 1B, and 7B parameters).
We take input and output token lengths of 160 and 140 on different GPUs.
The input and output token lengths are highly representative as they follow the median lengths in the ShareGPT dataset \cite{sharegpt}.

\begin{figure}[t]
	\centering
	\includegraphics[width=1\linewidth]{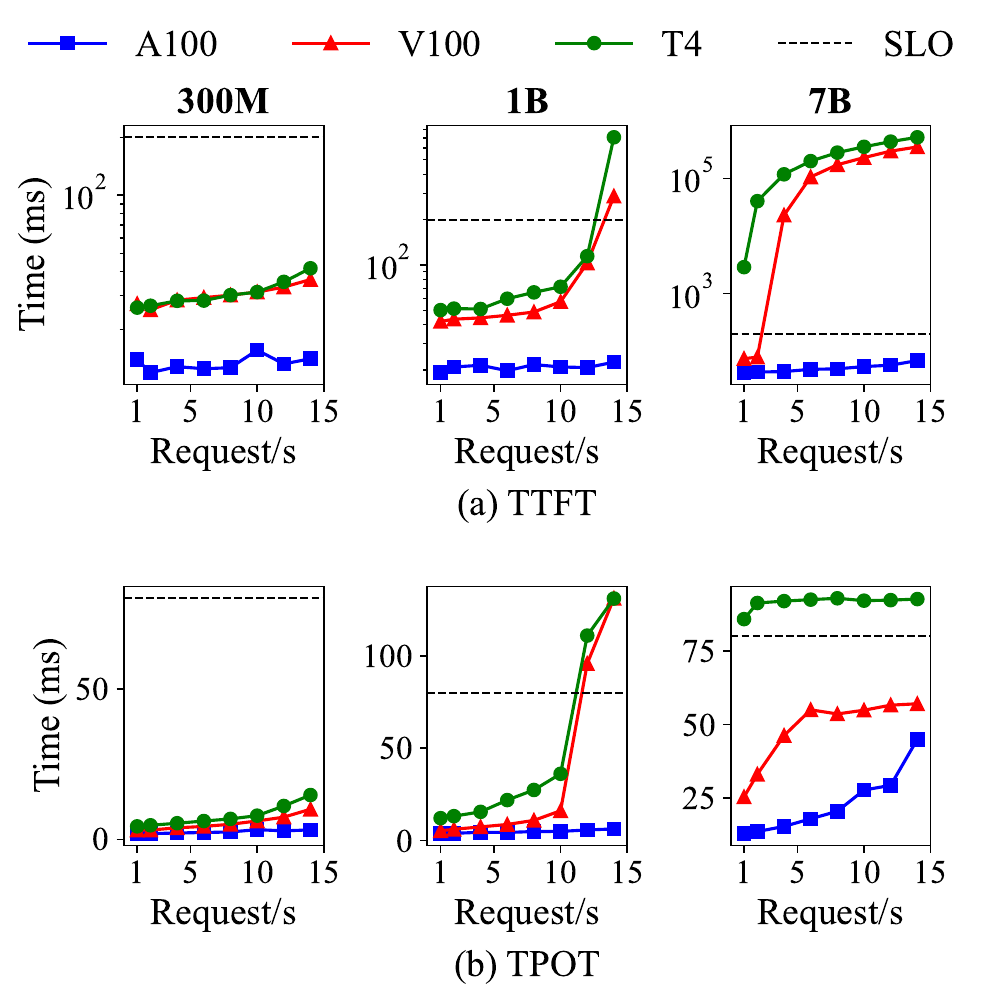}
	\caption{\textbf{TTFT (prefill) and TPOT (decoding) of different model sizes at different request rates on different GPUs.} The latency SLOs for prefill (TTF) and decoding (TPOT) phases are 200ms and 80ms, respectively.}\label{fig:m-latency-size}
\end{figure}

\paragraph{Latency.} \Cref{fig:m-latency-size} shows the latency of the prefill (TTFT) and decoding (TPOT) of different model sizes at different request rates across different GPU types as a function of request rate. The latency SLOs for TTFT and TPOT are 200ms and 80ms, respectively. The results highlight three key findings. 
First, as the model size increases, both TTPT and TPOT increase across all GPU types due to greater compute demands, leading to longer processing times. Second, larger model sizes reduce the request rate that can meet latency SLOs, with the largest models (7B) potentially failing to meet SLOs at any request rate. Third, at many request rates, older GPUs (V100 and T4) can meet latency SLOs comparable to newer GPUs (A100), particularly for smaller models, revealing opportunities for running smaller models using older GPUs effectively.  

\takeawaybox{
The prefill and decoding phases feature different characteristics, as prior works have shown~\cite{zhong2024distserve,patel2020clite}. 
Particularly for old GPUs, the prefill phase is more demanding as it is compute-bound, while the decoding phase is less demanding as it is memory-bound. 
Thus, disaggregating LLM serving enables using older GPUs for specific phases.
}

\begin{figure}[t]
	\centering
	\includegraphics[width=1\linewidth]{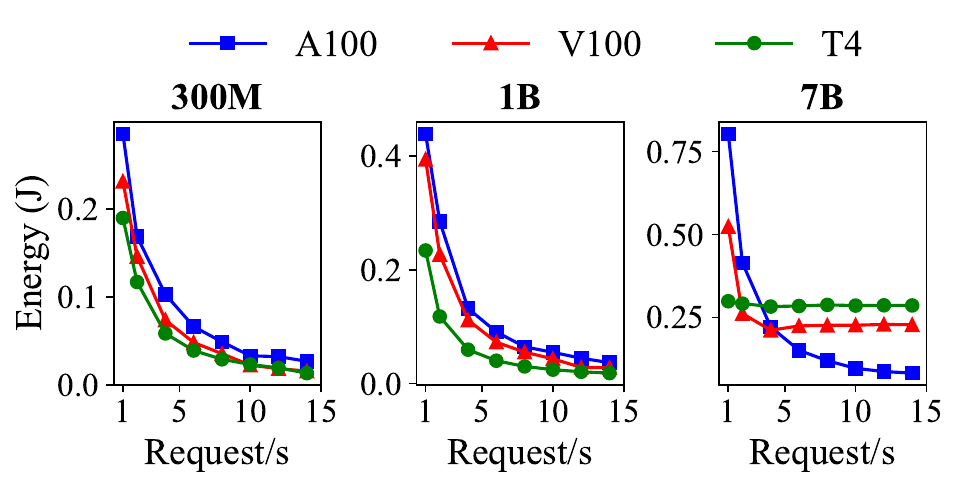}
	\caption{\textbf{Energy per token of different model sizes at different request rates on different GPUs.}  }\label{fig:m-energy-size}
\end{figure}

\paragraph{Energy.} \Cref{fig:m-energy-size} shows the energy consumption per token of different model sizes on different GPUs as a function of the request rate. The results highlight three key findings. First, energy increases with model size across all GPU types at the same request rate due to higher processing times and greater compute demands.  
Second, as request size grows, energy per token generally decreases. This is because larger requests improve GPU utilization, allowing the GPU to operate closer to its thermal design power (TDP). At this point, token throughput increases faster than power usage, stabilizing energy consumption per token when the GPU reaches optimal throughput within its TDP.  
Third, older GPUs (V100, T4) tend to be more energy-efficient than newer GPUs (A100) for smaller models (300M, 1B). However, this advantage diminishes for larger models like 7B, where older GPUs are only more energy-efficient at low request rates.  

\takeawaybox{
Small-size models can meet comparable latency SLOs as larger models at certain request rates, even on older GPUs, while providing significantly better energy efficiency. To meet latency SLOs, offloading computations to older GPUs is more effective in executing smaller models. For less complex LLM tasks, replacing large models on new GPUs with smaller ones on older GPUs can be more energy- and carbon-efficient.
}

{
\setlength{\belowcaptionskip}{-6pt}
\begin{figure}[!t]
	\begin{center}
		\includegraphics[width=.85\linewidth]{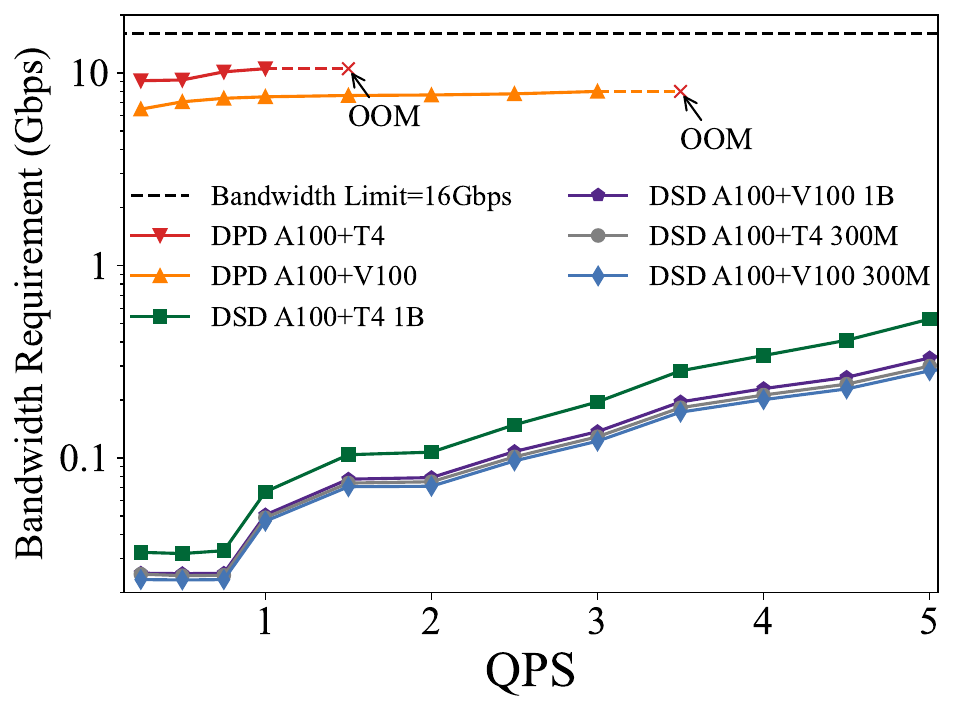}  \vspace{-0.1in}
	\end{center} %\vspace{-0.2in}
	\caption{\textbf{Bandwidth requirement comparisons between \DistPD{} (DPD) and \DistSD (DSD) when connecting A100 and T4/V100 at different request rates (QPS).} DPD experiments the 7B model only, with the prefill phase running on A100 and the decoding phase on T4/V100. For DSD, the 7B target model runs on A100, while the 1B or 300M draft model runs on T4/V100.}\label{fig:bandwidth}
\end{figure}
}

\subsection{Bandwidth Requirement Matters}\label{subsec:bandwidth}

As discussed in~\cref{subsec:related}, a key limitation of disaggregating the prefill and decoding phases is the high interconnect bandwidth requirement due to the need for transferring large KV caches. 
Although prior work like DistServe~\cite{zhong2024distserve} and Splitwise~\cite{patel2024splitwise} have discussed this issue, their solutions depend on high-end homogeneous or heterogeneous GPUs equipped with high-speed interconnects such as NVLink or InfiniBand. 
However, this limitation persists in scenarios where high-bandwidth connectivity is unavailable and becomes even more pronounced when data transfers occur between heterogeneous GPUs---where one is high-end and the other is older and lower-tier. 
In such cases, performance gains diminish and carbon reduction benefits are also reduced.  

\Cref{fig:bandwidth} illustrates the bandwidth requirements when connecting heterogeneous GPUs, such as A100 and T4/V100, with a 16Gbps bandwidth cap. The results show that when serving the LLaMA 7B model and transferring the KV cache between an A100 and T4/A100, the system quickly encounters out-of-memory (OOM) issues as QPS increases.  

To address this issue, we identify a second use case based on speculative decoding. In the experiments, the target model is 7B, while the draft models are smaller (1B and 300M). As shown in \Cref{fig:bandwidth}, \DistSD{} requires 65--434$\times$ lower bandwidth than \DistPD{} over a wide range of request rates since it only transfers tokens and corresponding probability distributions rather than the large KV caches.

\takeawaybox{
Disaggregating prefill and decoding phases on different GPUs is feasible only when high network connectivity is available. Otherwise, \DistSD{} is an alternative for reducing carbon emissions. Combining the results from~\cref{subsec:model-size}, running the draft model of speculative decoding on older GPU is viable and offers potential for high carbon efficiency.
}
\section{\SYSTEM{}}\label{sec:method}

\begin{figure}[!t]
	\begin{center}
		\includegraphics[width=0.75\linewidth]{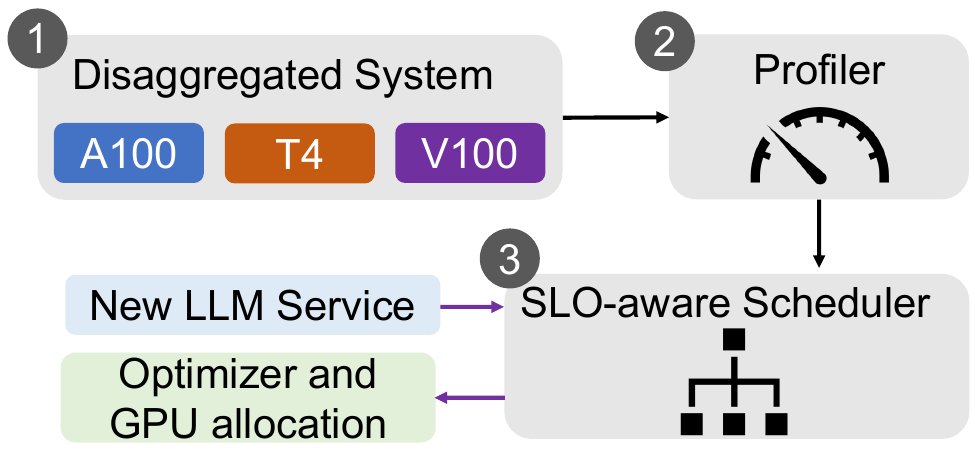}  \vspace{-0.1in}
	\end{center}\vspace{-0.1in}
	\caption{\textbf{Workflow of \SYSTEM{}.}}\label{fig:workflow}
\end{figure}

We present \SYSTEM{}, an SLO-aware LLM serving framework that minimizes carbon emissions using disaggregation on heterogeneous GPUs. \Cref{fig:workflow} shows the workflow of \SYSTEM{}, which consists of three components: \textcircled{1} a disaggregated system on heterogeneous GPUs, \textcircled{2} a profiler, and \textcircled{3} an SLO-aware scheduler. 
\textcircled{1} implements how we disaggregate two optimizers (\DistPD{} and \DistSD) on heterogeneous GPUs with minimal communication overhead. \textcircled{2} performs profiling to measure GPU performance and carbon emissions across request sizes and rates. \textcircled{3} selects the optimal optimizer and GPU configuration to minimize carbon emissions using profiled data. Next, we will elaborate on the design and implementation of each component.

\subsection{Disaggregation on Heterogeneous GPUs}

As \SYSTEM{} integrates two LLM serving optimizers (\DistPD{} and \DistSD{}), each optimizer disaggregates different execution components onto different GPUs. Consequently, their disaggregation designs differ. We will elaborate on each design as follows.

\noindent \textbf{\DistPD{}.} Because the decoding phase is less compute-intensive than the prefill, \DistPD{} allocates the prefill phase to newer GPU and the decoding phase to older GPU while meeting latency SLOs. 
\DistPD{} differs from prior disaggregation work like DistServe~\cite{zhong2024distserve} and Splitwise~\cite{patel2024splitwise} in two ways. First, prior work disaggregated both phases on high-end GPUs (e.g, H100, A100), neglecting the potential of reusing old, low-performing GPUs. Second, while prior work prioritized performance and cost optimization, \SYSTEM{} focuses on carbon reduction. Although we identify \DistPD{} as a use case for reusing old GPUs, as indicated in~\cref{subsec:bandwidth}, the high network bandwidth requirement of this disaggregation strategy limits its applicability in scenarios with limited network connectivity, leading to our second optimizer introduced below.

\begin{figure}[!t]
	\begin{center}
		\includegraphics[width=\linewidth]{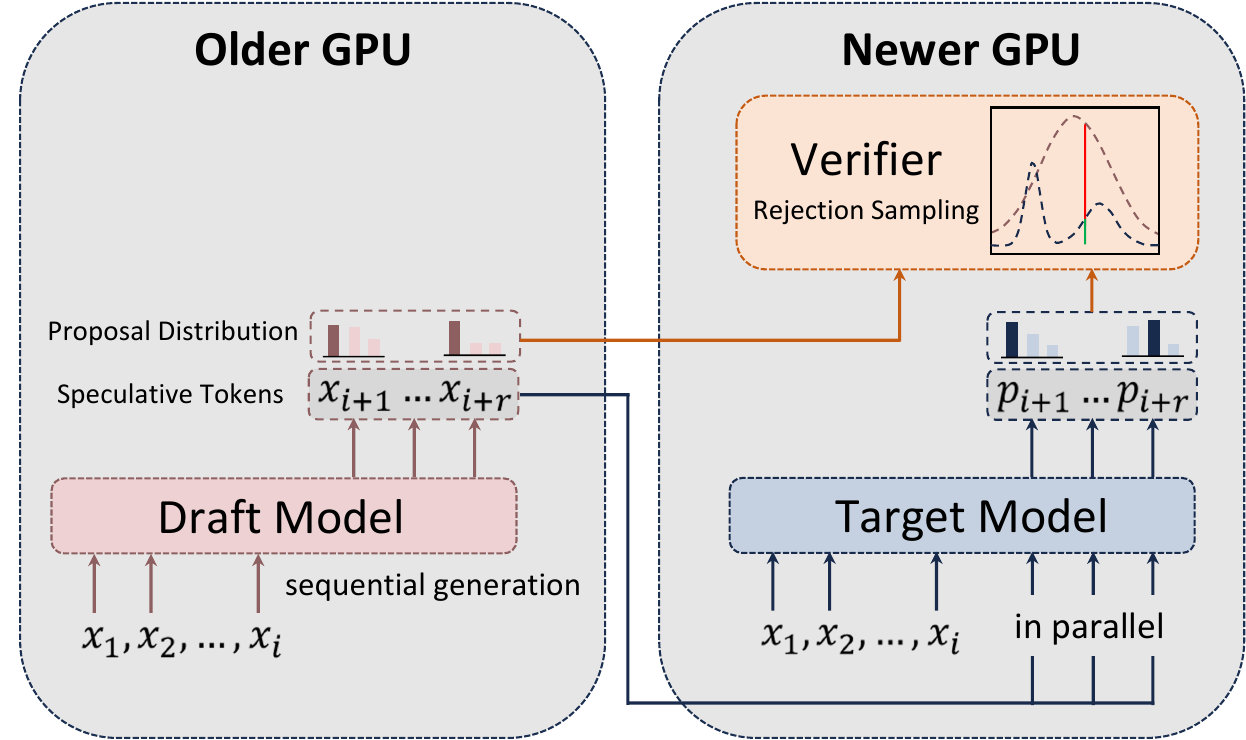}  \vspace{-0.1in}
	\end{center}\vspace{-0.3in}
	\caption{\textbf{Disaggregating draft and target models on heterogeneous GPUs for speculative decoding in \SYSTEM{}.}}\label{fig:spec-dist}
\end{figure}

\begin{figure}[!t]
	\begin{center}
		\includegraphics[width=\linewidth]{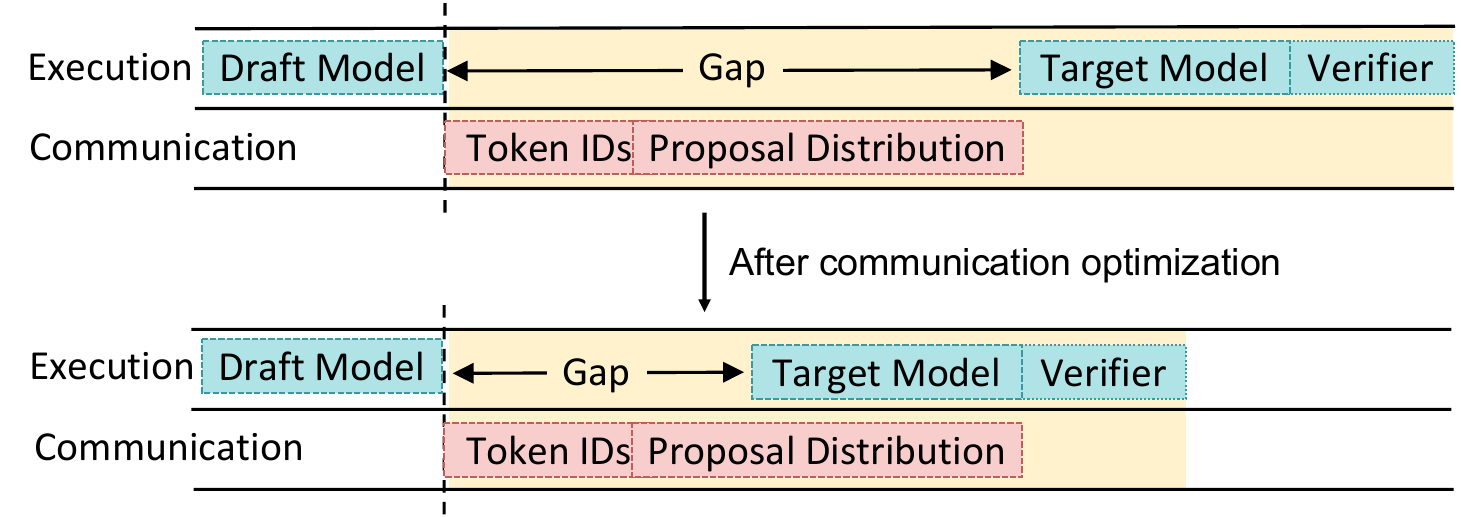}  \vspace{-0.1in}
	\end{center}\vspace{-0.3in}
	\caption{\textbf{Communication overlapping optimization for \DistSD{}.} Green blocks indicate the execution of each component. Red blocks indicate data transfer. The yellow background indicates the time comparisons. }\label{fig:spec-opt}
\end{figure}

\noindent \textbf{\DistSD{}.} As discussed in~\cref{subsec:model-size}, smaller models can achieve SLOs comparable to larger models, even on older GPUs, while providing better energy efficiency. Based on this observation, \DistSD{} assigns the draft model to an older GPU and the target model (along with the verifier) to a newer GPU, as illustrated in~\Cref{fig:spec-dist}, thus reducing energy consumption and carbon emissions. However, separating the draft and target models across GPUs inevitably introduces communication overhead. 

To maximize carbon savings while minimizing this overhead, we analyze the temporal dependencies between the two models to explore the opportunity for communication optimization. 
We observe that the speculative token IDs and the draft model's probabilities differ significantly in size, with the probabilities being $V$ times larger, where $V$ is the vocabulary size. Importantly, the verifier does not require the draft probabilities until after the target model has processed the speculative tokens.  

Based on this insight, we illustrate the communication optimization process in~\Cref{fig:spec-opt}. We first transfer only the speculative token IDs, which are relatively small in size. While the target model performs decoding using these token IDs, we overlap communication and computation by asynchronously transferring the larger draft probabilities in parallel. This approach minimizes communication overhead by leveraging the delayed dependency of probabilities, effectively hiding data transfer delays between the draft and target models.

\subsection{Profiler}

\SYSTEM{} includes a profiler to collect the latency, energy consumption, and carbon emissions for different GPUs across a range of LLM serving workloads (e.g., varying request sizes and rates). For \DistPD{}, it collects data for the prefill and decoding phases separately running on different types of GPUs. For \DistSD{}, it collects data for different model sizes running on different types of GPUs. This profiling data will be used as the database for the SLO-aware scheduler described below.

\subsection{SLO-Aware Scheduler}

\begin{algorithm}[t]
\caption{The SLO-aware Scheduler}\label{alg:search}
\KwIn{Profiling database $\mathbf{D}$, serving workload characteristics $\mathbf{W}=\{(req\_size,qps)\}$, $SLO_{target}$, priority}
\KwOut{Optimal configuration $\mathbf{Optimal}$ for each workload}
\BlankLine
\SetKwFunction{CF}{CollaborativeFiltering}
\SetKwFunction{FindOptimal}{FindOptimalConfiguration}
\SetKwFunction{Fallback}{FallbackStrategy}
% \textbf{Step 1: Matrix Completion} \\
$\mathbf{C}, \mathbf{SLO_{att}} \gets$ \CF($\mathbf{D}$)
% \textbf{Step 2: Identify Optimal Configurations Under Constraints} \\
\ForEach{$j \in \mathbf{W}$}{
    $\mathbf{Feasible}_j \gets \{i \mid \mathbf{SLO_{att}}[i,j] \geq SLO_{target}\}$ \\
    \eIf{$\mathbf{Feasible}_j \neq \emptyset$}{
        $\mathbf{Optimal}[j] \gets \arg \min_{i \in \mathbf{Feasible}_j} \mathbf{C}[i,j]$ \\
    }{
        $\mathbf{Optimal}[j] \gets$ \Fallback($\mathbf{SLO_{att}}, \mathbf{C}, j, \text{priority}$)
    }
}
\Return{$\mathbf{Optimal}$}
\BlankLine
\SetKwBlock{Procedure}{Procedure \Fallback{}}{}
\Procedure{
    \uIf{priority = 'SLO'}{
        \Return{$\arg \max_{i} \mathbf{SLO\_Att}[i,j]$}
    }
    \Else{
        \Return{Default configuration}
    }
}
\end{algorithm}

{
\setlength{\belowcaptionskip}{-3pt}
\setlength{\textfloatsep}{3pt}
\begin{figure}[!t]
	\begin{center}
		\includegraphics[width=0.7\linewidth]{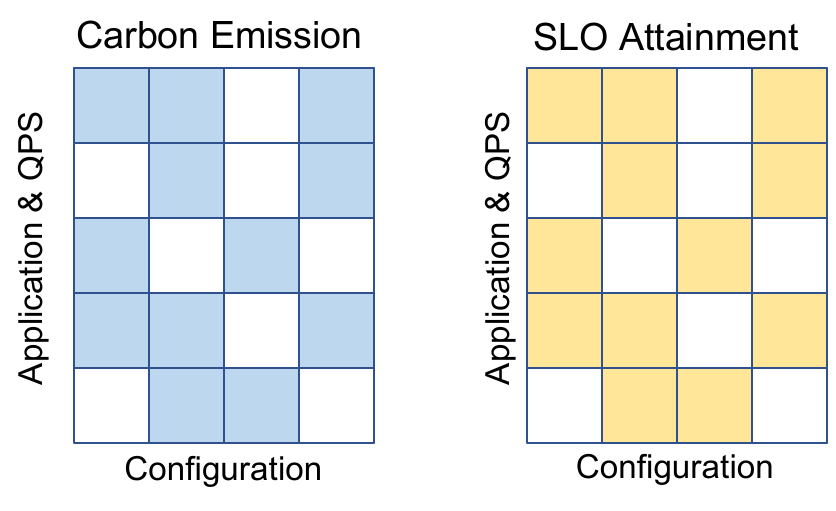}  \vspace{-0.1in}
	\end{center}\vspace{-0.1in}
	\caption{\textbf{Matrix formulation for carbon emissions and SLO attainment.} Shaded regions represent known data, while blank entries represent missing data. Each entry represents the carbon emission or SLO attainment for a specific application, QPS level, and configuration. \SYSTEM{} uses collaborative filtering to fill in the missing entries. }\label{fig:matrix}
\end{figure}
}

\SYSTEM{} incorporates an SLO-aware scheduler to identify the optimal \emph{configuration}---the combination of an LLM serving optimizer and its GPU allocation---to execute a given LLM service workload with specified request size and query rate (queries per second or QPS). Before initiating the search, missing latency and carbon emission values for unseen request sizes and rates must be estimated.  

To address the search across both workload and configuration dimensions, \SYSTEM{} organizes the search space into two two-dimensional matrices: one matrix, \(\mathbf{C}\), representing carbon emissions, and another, \(\mathbf{SLO_{att}}\), representing SLO attainment. In both matrices, the row represents combinations of applications and QPS levels, and the column represents configurations. The entries in these matrices are carbon emissions and SLO attainment, respectively. \Cref{fig:matrix} illustrates the two matrices with missing values. To fill in missing values in these matrices, \SYSTEM{} employs collaborative filtering, a technique previously used in resource management~\cite{delimitrou2013paragon,ding2019generative}. This process is performed only once after profiling.  

Once the matrices are complete, \SYSTEM{} searches for the configuration that minimizes carbon emissions while meeting the SLO. If no such configuration exists, a fallback strategy is applied based on the specified priority, such as maximizing SLO attainment or selecting a default configuration The search algorithm is summarized in~\Cref{alg:search}.

% \tys{This scheduler determines the optimal system configuration given the workload characteristics, i.e., the request size and QPS. It uses collaborative filtering (CF) for matrix completion as in Paragon~\cite{delimitrou2013paragon} to address missing performance and carbon metrics in the profiling data, avoiding exhaustive search in the whole configuration space. Here the rows of matrices $\mathbf{C}$, $\mathbf{SLO_{att}}$ are applications and QPS levels, the columns are system configurations, and the values are carbon footprints and SLO attainment, respectively. Matrix completion only needs to be done once after the offline profiling stage. Then the algorithm identifies feasible configurations under SLO constraints and select the one with the minimum carbon footprint. If no configuration meets the SLO constraint, a fallback strategy is employed based on the specified priority (e.g., maximizing SLO attainment or selecting a default configuration).}

% \SYSTEM{} selects optimizer---\DistPD{} or \DistSD{}---and the GPU allocation based on its SLO-aware online scheduler. Given an LLM service with a specified latency SLO, \SYSTEM{} searches the space of optimizers and GPU allocations to identify the combination that minimizes total carbon emissions while meeting the latency SLO. For online LLM services, \SYSTEM{} dynamically updates the optimizer and GPU allocation as workloads and SLOs change.

\section{Carbon Efficiency Analysis}\label{sec:analysis}

This section delves into the carbon efficiency analysis of disaggregation, comparing scenarios with both new and old GPUs to that using only new GPU. We aim to answer three questions: 1) When does disaggregation yield greater carbon savings? 2) How does carbon intensity influence carbon savings? 3) How does GPU lifetime influence carbon savings?

Consider two cases that run the same LLM service. 
\begin{enumerate}[label=\textbf{Case \arabic*}:]
    \item \textbf{(Standalone)} The LLM service runs on a new GPU $A$, incurring execution time $t_A$, energy consumption $N_A$, operational carbon emission $O_A$, and embodied carbon emission $E_A$.
    \item \textbf{(Disaggregation)} The same LLM service runs on a new GPU $A$ and an old GPU $B$, incurring execution time $t_A'$, $t_B$, energy consumption $N_A'$, $N_B$, operational carbon emission $O_A'$, $O_B$, and embodied carbon emission $E_A'$, $E_B$. 
\end{enumerate}
Our analysis will be based on the following assumptions:
\begin{enumerate}[label=\textbf{A.\arabic*}:]
    \item Both cases use electricity from the same power grid with carbon intensity $\alpha$.
    \item The carbon emissions of communication overhead between two GPUs in Case 2 are negligible.
    \item The time difference between running $A$ in Case 1 and Case 2 is substantially shorter than the time to run $B$ in Case 2. This assumption is practical, as offloading computations to older GPUs typically incurs longer running times as well as higher total embodied carbon than running the same computations on newer GPUs.
\end{enumerate}

\noindent\textbf{When does disaggregation yield greater carbon savings?} 
To reduce total carbon emissions in Case 2, we need
\begin{align}
& O_A + E_A > O_A' + E_A' + O_B + E_B  \nonumber \\
\Rightarrow  &  O_A - (O_A' + O_B)  > (E_A' + E_B) - E_A  \nonumber
\end{align}
Due to A.3, $(E_A' + E_B) - E_A >0$, as disaggregation, involving two GPUs, typically incurs higher embodied carbon than standalone. Therefore, we have 
\begin{align}
& O_A - (O_A' + O_B) > 0   \nonumber \\
\Rightarrow  &  O_A > O_A' + O_B  \nonumber \\
\Rightarrow  & E_A \cdot \alpha > ( E_A' + E_B) \cdot \alpha  \nonumber \\
\Rightarrow  & E_A  >  E_A' + E_B \label{eq:q1}
\end{align}
\conclusionbox{
\Cref{eq:q1} indicates that disaggregated systems must consume less energy than standalone systems to achieve carbon savings. 
}

\noindent\textbf{How does carbon intensity influence carbon savings?} To quantify carbon savings, we can calculate
\begin{align}
\frac{O_A' + E_A' + O_B + E_B}{O_A + E_A} &= \frac{(N_A'+ N_B)\cdot \alpha + E_A'  + E_B}{N_A \cdot \alpha + E_A}  \nonumber \\
&=
\frac{\frac{N_A'+ N_B}{N_A }\left(N_A \cdot \alpha + E_A - E_A\right) +  E_A'  + E_B}{N_A \cdot \alpha + E_A} \nonumber \\
&=\frac{N_A'+ N_B}{N_A } + \frac{ E_A'  + E_B - \frac{N_A'+ N_B}{N_A }E_A}{N_A \cdot \alpha + E_A}  \label{eq:q2}
\end{align}

\conclusionbox{
\Cref{eq:q2} indicates that carbon savings increase with increasing carbon intensity $\alpha$ and decrease with decreasing $\alpha$.
}

\noindent\textbf{Hoes does GPU lifetime influence carbon savings?}  Let $\mathcal{A}$ and $\mathcal{B}$ be the total embodied carbon emissions of GPU $A$ and $B$, respectively. $T_A$ and $T_B$ are the lifetimes of the GPUs $A$ and $B$, respectively. We keep deriving \Cref{eq:q2} and simplify it by removing constants, and have
\begin{align}
\frac{O_A' + E_A' + O_B + E_B}{O_A + E_A} &\approx  
\frac{ E_A'  + E_B - \frac{N_A'+ N_B}{N_A }E_A}{N_A \cdot \alpha + E_A}   \nonumber \\
&=  \frac{ \frac{t_A'}{T_A}\mathcal{A}  + \frac{t_B}{T_B}\mathcal{B} - \frac{N_A'+ N_B}{N_A }\frac{t_A}{T_A}\mathcal{A}}   {N_A \cdot \alpha +\frac{t_A'}{T_A}\mathcal{A} }  \nonumber \\
&=  \frac{ \left( \frac{t_A'}{T_A} -  \frac{N_A'+ N_B}{N_A }\frac{t_A}{T_A} \right) \mathcal{A}  + \frac{t_B}{T_B}\mathcal{B}}   {N_A \cdot \alpha +\frac{t_A'}{T_A}\mathcal{A} }   \nonumber \\
 &\approx  \texttt{Const.} +  \frac{ \frac{t_B}{T_B}\mathcal{B}}   {N_A \cdot \alpha +\frac{t_A'}{T_A}\mathcal{A} } \label{eq:q3}
\end{align}

When fixing $T_B$, decreasing $T_A$ will increase carbon savings. When fixing $T_A$, increasing $T_B$ will increase carbon savings.

\conclusionbox{
\Cref{eq:q3} indicates disaggregation yields maximum carbon savings when combined with newer GPUs that have been in service for a short time and older GPUs that have been operating for a long time.
}

In summary, disaggregation systems achieve the highest carbon savings under high carbon intensity, high energy efficiency, and a combination of shorter service time for newer GPUs and longer service time for older GPUs.

\section{Implementation}

\begin{table*}[t]
    \centering
    \caption{\textbf{Datasets for evaluation and SLO requirements.} }
    \begin{tabular}{lllllll}
    \toprule
     Dataset & Task & TTFT SLO & TPOT SLO & P25 Req. Size  & P50 Req. Size  & P75 Req. Size \\
    \midrule
        ShareGPT~\cite{sharegpt} & Chatbot & 200ms & 80ms & (24,24) & (160,140) & (510,357) \\
        HumanEval~\cite{Chen2021EvaluatingLL} & Code Completion & 125ms & 200ms & (108,31) & (136,55) & (182,88) \\
        LongBench~\cite{Bai2023LongBenchAB} & Summarization & 15s & 150ms & (1134,201) & (1495,275) & (1817,352)\\   
        % Tianyao: removed requests with output length < 50 and context length > 2k in longbench
        % Stats does not match our experiment settings, need to rerun after OSDI deadline
    \bottomrule
    \end{tabular}    
    \label{tab:workload_stats}
\end{table*}

We implemented \SYSTEM{} on top of the open-source LLM serving platform vLLM~\cite{kwon2023efficient}, integrating the profiler and \DistSD{} method as plug-ins. For the \DistPD{} method, we leveraged community implementations of vLLM based on the design principles of Splitwise~\cite{patel2024splitwise}.  

% ### Profiler Implementation  
The profiler uses \verb|pynvml| APIs to collect GPU power consumption data with minimal overhead. Communication and profiling data aggregation across multiple servers are managed via RESTful APIs, ensuring robustness and efficiency in distributed environments. The profiler is implemented in 1.7K lines of Python and Linux Shell code.

% ### \DistSD{} Implementation  
The \DistSD{} method was implemented on top of vLLM with an additional 1K lines of Python code. Following vLLM's architecture, we utilized Ray actors to create GPU workers for the draft model inference, assigning older-generation GPUs in a heterogeneous cluster for running the draft model workers. Unlike the original vLLM design, we introduced two distinct parallelism groups on heterogeneous nodes to support parallel inference. One group was responsible for the draft model, while the other handled the primary serving instance, including the serving API, target model, and verifier. To facilitate communication between these groups, we employed NCCL along with PyTorch’s asynchronous communication primitives, \textit{isend} and \textit{irecv}, to avoid blocking GPU computations during data transmission.

% \tys{We implemnet our DistCarbon framework on top of the open-source LLM serving framework vLLM~\cite{kwon2023efficient}, where the profiler and the Dist-Spec-Decode method works as plug-ins of vLLM.
% The profiler leverages} \verb|pynvml| \tys{APIs to collect GPU power consumption data with minimal overhead. 
% We use RESTful APIs to coordinate the communication and gather profiling data from multiple servers, ensuring robustness and efficient data aggregation across distributed environments. 
% The profiler is implemented with 1.7K lines of Python and Linux Shell code.
% For the Dist-Prefill-Decode method, we use community implementations of vLLM following the design of Splitwise~\cite{patel2024splitwise}.}
% \yrw{We implemented the Dist-Spec-Decode method on top of the open-source LLM serving framework vLLM, with 1K lines of Python code. Building on vLLM's design, we used Ray actors to create GPU workers for handling draft model inference. Our setup utilized a heterogeneous GPU cluster, assigning older-generation GPUs to run the draft model workers. Unlike the original design, we established two separate parallelism groups on the heterogeneous nodes, enabling parallel inference within each group. One group was dedicated to the draft model, while the other handled the main serving instance, which included the serving API, target model, and verifier. For communication between the groups, we used NCCL and employed PyTorch’s asynchronous communication primitives, \textit{isend} and \textit{irecv}, to prevent blocking GPU computations during data transmission. }

\section{Evaluation} \label{sec:evaluation}

We present a comprehensive evaluation as follows. First, we show the main results of carbon emission savings achieved by \SYSTEM{} under various workloads (\cref{subsec:res-carbon-saving}). Second, we present the latency results of \SYSTEM{} (\cref{subsec:res-latency}). We then show the impact of network bandwidth on disaggregation and carbon savings (\cref{subsec:res-bandwidth}), followed by sensitivity analyses of varying carbon intensity (\cref{subsec:res-ci}) and GPU lifetime (\cref{subsec:res-lifetime}).

\subsection{Methodology}

In this section, we discuss the system setup, models and configurations, workloads, measurement approach, and metrics. 

\noindent\textbf{System setup.}
We evaluate \SYSTEM{} using Google Cloud Platform (GCP) servers: machine type \texttt{a2-highgpu-1g} with one A100 GPU, \texttt{n1-standard-8} with one T4/V100 GPU. 
When serving in a disaggregated mode, servers are connected within the GCP network with a default bandwidth of 16Gbps.  

\noindent\textbf{Models and configurations.}
We evaluate \SYSTEM{} with Llama models of varying sizes: 7B, 1B, and 300M parameters~\cite{meta_llama}. \SYSTEM{} incorporates the following configurations: 

\begin{itemize}
    \item \textbf{Standalone}: serve a 7B Llama model on a single A100. 
    \item \textbf{SpecDecode} (Single-GPU Speculative Decoding): serve target (7B) and draft (1B or 300M) models from speculative decoding on a single A100.  
    \item \textbf{DPD} (\DistPD{}): serve prefill and decoding phases of a 7B Llama model on different GPUs: the prefill stage on newer GPU (A100), and the decoding stage on older GPUs (T4 or V100).
    \item \textbf{DSD} (\DistSD{}): use 1B and 300M Llama models as choices of draft model running on old GPUs (T4 or V100), and 7B Llama model as the target model on newer GPU (A100).
\end{itemize}
\SYSTEM{} automatically selects these configurations to minimize carbon emissions and meet latency SLOs. 
We take the Standalone A100 setup as \emph{baseline}, representing a system that does not reuse old GPUs but only uses newer GPUs.

% We use a 7B-parameter version for single-GPU evaluation, and disaggregated prefill and decoding. 
% In speculative decoding, the same 7B version is used as the target model, and a 1B-parameter and a 300M-parameter version are two alternative draft models. 

\noindent\textbf{Workloads.}
We evaluated the same LLM serving datasets as in previous work DistServe~\cite{zhong2024distserve}: ShareGPT~\cite{sharegpt}, HumanEval~\cite{Chen2021EvaluatingLL}, and LongBench~\cite{Bai2023LongBenchAB}. They represent three types of LLM applications: ChatGPT-like chatbot, code completion, and summarization, respectively. 
\Cref{tab:workload_stats} summarizes the SLOs, and input and output distributions for each workload.  
% HumanEval and LongBench have relatively narrow distributions of input/output token numbers, whereas ShareGPT has a more widespread distribution. 
% Therefore, we take the median input/output token numbers for HumanEval and LongBench, and the three 
% quantiles (\nth{25}, \nth{50}, and \nth{75}) for ShareGPT. 
% \todo{Explain why we only evaluate different sizes for shareGPT.}
% \tys{The 25th output size in the data we used is only 7 because we used context from QA tasks!}
When evaluating a fixed input/output token length for a workload, we truncate the prompts to the specific input length and limit the output length. This way, the carbon emission and performance results of different prompts are comparable. 
{
\setlength{\belowcaptionskip}{-7pt}
\begin{figure}[t]
    \centering
    \includegraphics[width=\linewidth]{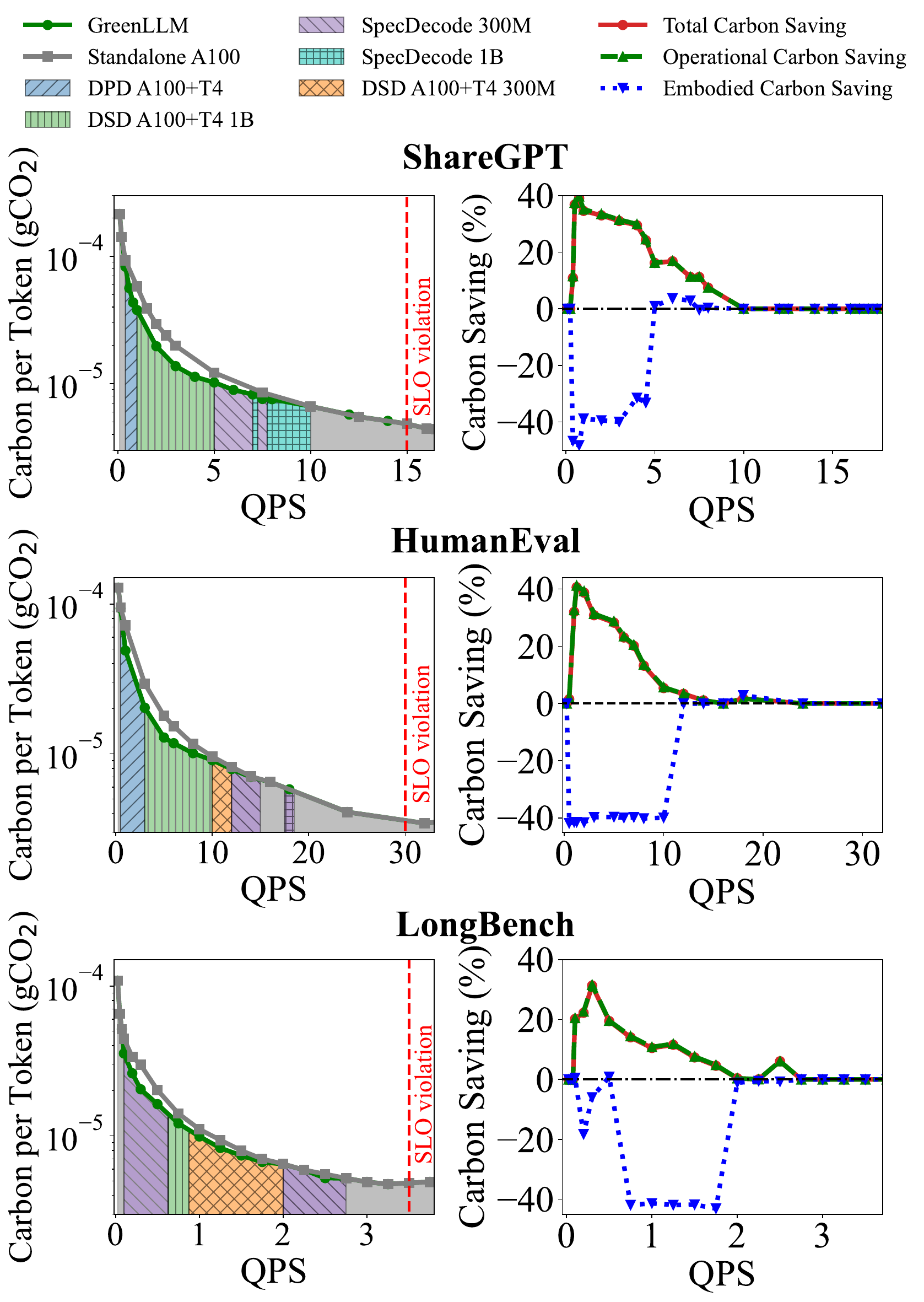}
    \caption{\textbf{Carbon per token and carbon savings comparison between \SYSTEM{} and the baseline Standalone A100.} The left shows \SYSTEM{}'s optimal configurations at various QPS. The right shows the corresponding total carbon savings breakdown into operational and embodied emissions.
    }\label{fig:E-carbon-main}
\end{figure}
}

\begin{figure*}[t]
    \centering
    \includegraphics[width=.85\linewidth]{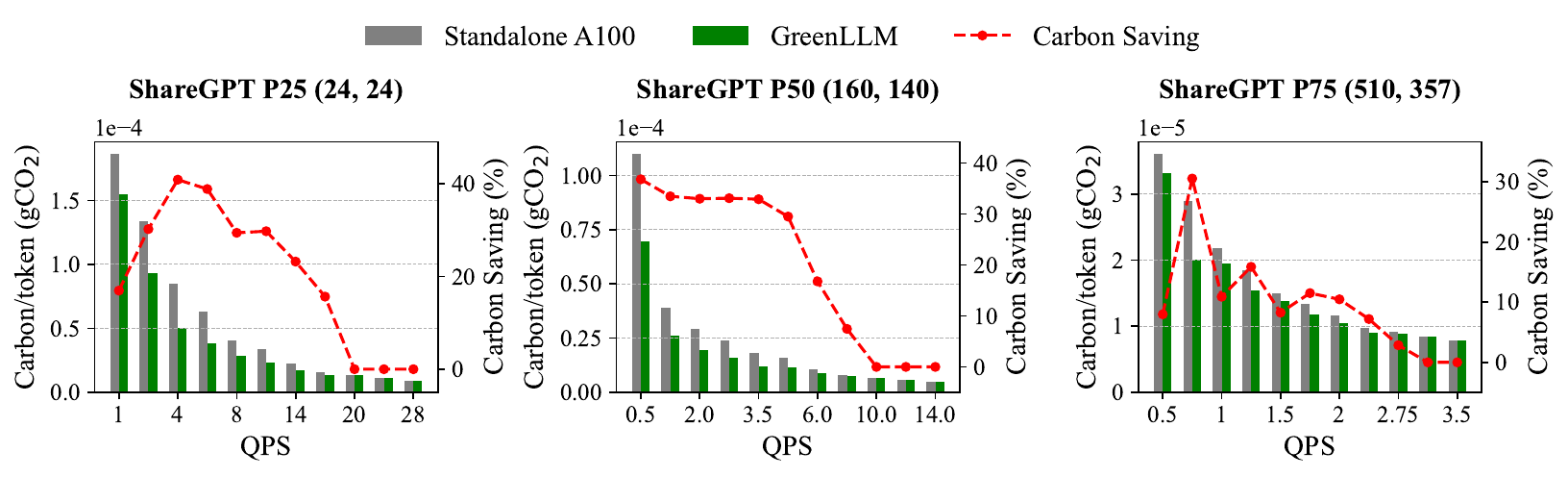}
    \caption{\textbf{Carbon emission per token and Savings at different request sizes on the ShareGPT dataset.}}
    \label{fig:ev-carbon-sharegpt-sizes}
\end{figure*}

\begin{figure*}[t]
    \centering
    \includegraphics[width=\linewidth]{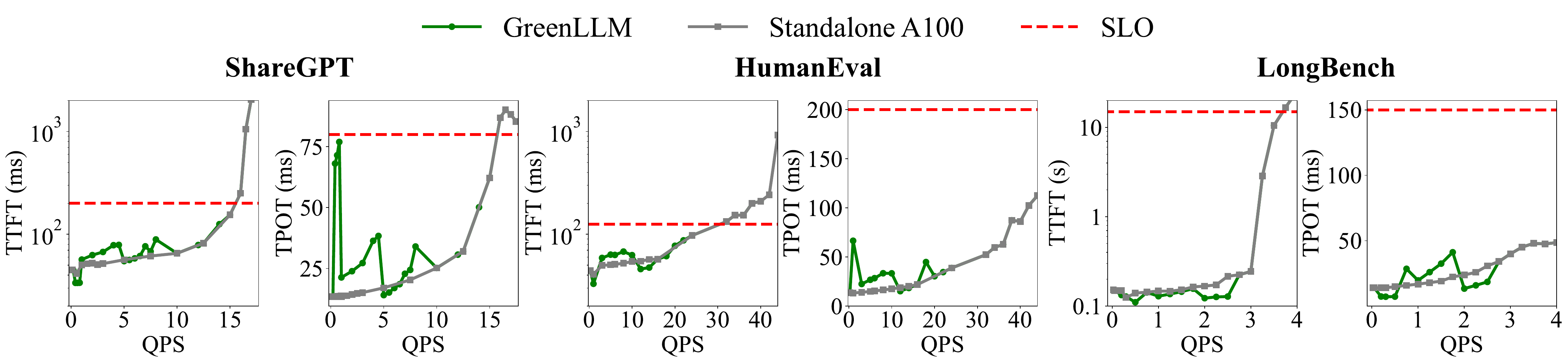}
    \caption{\textbf{Latency (TTFT and TPOT) at various QPS across three datasets.}}\label{fig:E-latency-main}
\end{figure*}

\noindent\textbf{Measurement and metrics.}
We measure execution time by inserting a timer in \SYSTEM{} and record token generation timestamps. During model execution, we measure the GPU power consumption every 200 ms using the \texttt{pynvml} library. Carbon emissions are calculated using the carbon intensity of 261 gCO$_{2}$/kWh from the CISO grid by default.

We evaluate \SYSTEM{} using the following metrics:

\begin{itemize}
    \item Carbon Per Token: gCO\textsubscript{2} of carbon emission per token. We model carbon emissions based on the carbon accounting method in~\Cref{subsec:carbon-account}. 
    \item Time To First Token (TTFT): the latency between prompt reception and the generation of the first token. 
    \item Time Per Output Token (TPOT): the latency between two consecutive tokens. 
\end{itemize}

We evaluate these metrics by varying the query per second (QPS) and input/out sizes conditions to holistically assess the carbon efficiency of \SYSTEM{}.

\subsection{Carbon Emission Savings}\label{subsec:res-carbon-saving}

In this section, we evaluate \SYSTEM{}'s carbon emissions and savings. For each workload, we assess carbon emissions and savings under varying QPS and input/output sizes. We also analyze the breakdown of embodied and operational carbon emissions improvement.

\noindent\textbf{Variable QPS.}
\Cref{fig:E-carbon-main} shows the carbon emission result, where the left column shows the carbon per token under different QPS values while the right column shows the corresponding carbon emission savings per token compared to the standalone A100 baseline. 
The input and output sizes of requests follow the median value of each dataset. 
Overall, \SYSTEM{} can effectively save up to 31.3--40.6\% carbon emissions when SLO attainment is 90\% (i.e., meeting SLO 90\% of the time). %\todo{please use precise numbers -- peak from the red curves. }
Carbon emission savings occur when the QPS lies in the lower range of each workload. 
In particular, we notice that QPS ranges that lead to the highest carbon emission savings vary among the three workloads. 
LongBench features the highest input/output token numbers, therefore \SYSTEM{} benefits from older GPUs only under lower QPS range [0.75,2] %(\todo{x--y}); 
HumanEval is on the contrary as it has lower input/output tokens, and thus old GPUs can provide benefits across a wider QPS range [0.5,11] %(\todo{x--y}). 
% \todo{I want to explain the choice of optimal configs under each QPS range. But the results look random?}

\noindent\textbf{Variable input/output sizes.}
We next investigate the carbon savings upon different input/output sizes.
ShareGPT \cite{sharegpt} is a highly representative dataset as it is based on prompts from real-world users. Therefore, we further evaluate token lengths at the three percentiles: P25, P50, and P75.
% following the methodology in \Cref{sec:method}.
\Cref{fig:ev-carbon-sharegpt-sizes} the per-token carbon emissions and carbon savings from \SYSTEM{} under these lengths.
Overall, under the same QPS, as the input/output sizes increase, the per-token carbon emission goes down as the carbon emissions are amortized among the tokens. 
At the same time, as the size increases, the QPS range that yields carbon savings becomes lower. This is because larger input sizes have higher computation loads and cause older GPUs to be less efficient. 
%\todo{how to explain the ups and downs in the third sub-figure in figure 10}

\noindent\textbf{Carbon savings breakdown.}  To further analyze the specific components of \SYSTEM{}'s carbon savings, we show the breakdown of the savings from operational and embodied carbon in the second column of Figure~\ref{fig:E-carbon-main}. At lower QPS, the primary source of carbon savings stems from reductions in operational carbon within disaggregated modes. This is because using older GPUs in disaggregated modes increases embodied carbon. As QPS increases, \SYSTEM{} prioritizes the SpecDecode method on a standalone A100, resulting in significant reductions in embodied carbon. This shift is driven by the fact that at higher QPS, using smaller draft models to speculatively generate output tokens leads to substantial improvements in TPOT, ultimately resulting in lower end-to-end latency. As indicated by Equation~\ref{eq:carbon-eb}, lower latency translates to reduced embodied carbon. In conclusion, this experiment demonstrates that \SYSTEM{} can effectively reduce both operational and embodied carbon by selecting the optimal configuration across a wide range of QPS.

\subsection{Latency}\label{subsec:res-latency}

In this section, we evaluate the latency of \SYSTEM{} and its SLO attainment under variable QPS.

\noindent\textbf{Variable QPS.} \Cref{fig:E-latency-main} shows the TTFT and TPOP latencies of \SYSTEM{} and the standalone A100 baseline under variable QPS. 
\SYSTEM{} automatically reconfigures the GPUs to meet SLO (the red line). Therefore, \SYSTEM{} meets both the TTFT and TPOP SLOs unless the QPS exceeds the capability of the standalone A100. 
In lower QPS ranges, \SYSTEM{} can experience a higher latency than standalone A100 because of old GPU reuse but still remains below SLO. 
On the other hand, \Cref{fig:E-carbon-main} demonstrates substantial carbon emission reductions during these SLO ranges, indicating that the \SYSTEM{} is effective at reconfiguring the GPUs to minimize carbon emissions while meeting SLOs.

\begin{figure}[t]
    \centering
    \includegraphics[width=\linewidth]{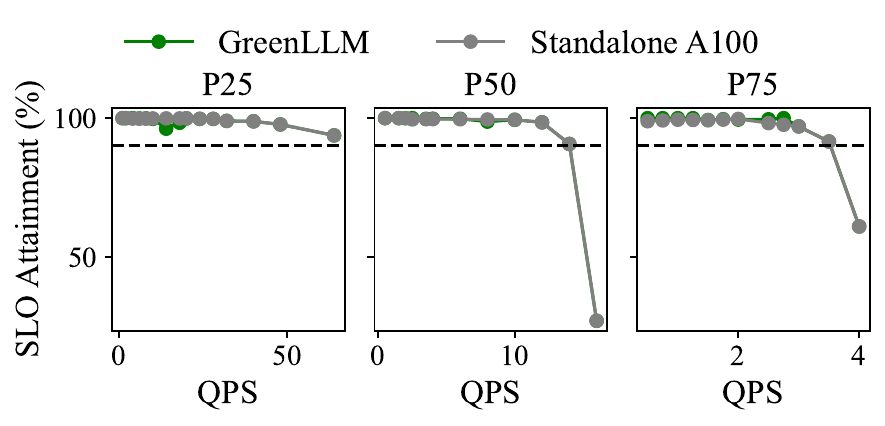}
    \caption{\textbf{SLO attainment at three different request sizes on the ShareGPT dataset.} The black dashed horizontal line indicates the SLO attainment of 90\%.}\label{fig:E-slo-sizes}
\end{figure}

\noindent\textbf{SLO attainment.} Figure~\ref{fig:E-slo-sizes} compares the SLO attainment of \SYSTEM{} and a standalone A100 configuration for three different request sizes (P25, P50, and P75) from the ShareGPT dataset. The black dashed horizontal line indicates the 90\% SLO threshold. The results show that \SYSTEM{} can effectively handle different request sizes while maintaining SLO performance similar to the standalone A100, but with the added benefit of greater carbon savings (\Cref{fig:ev-carbon-sharegpt-sizes}). 

\begin{figure}[t]
    \centering
    \includegraphics[width=\linewidth]{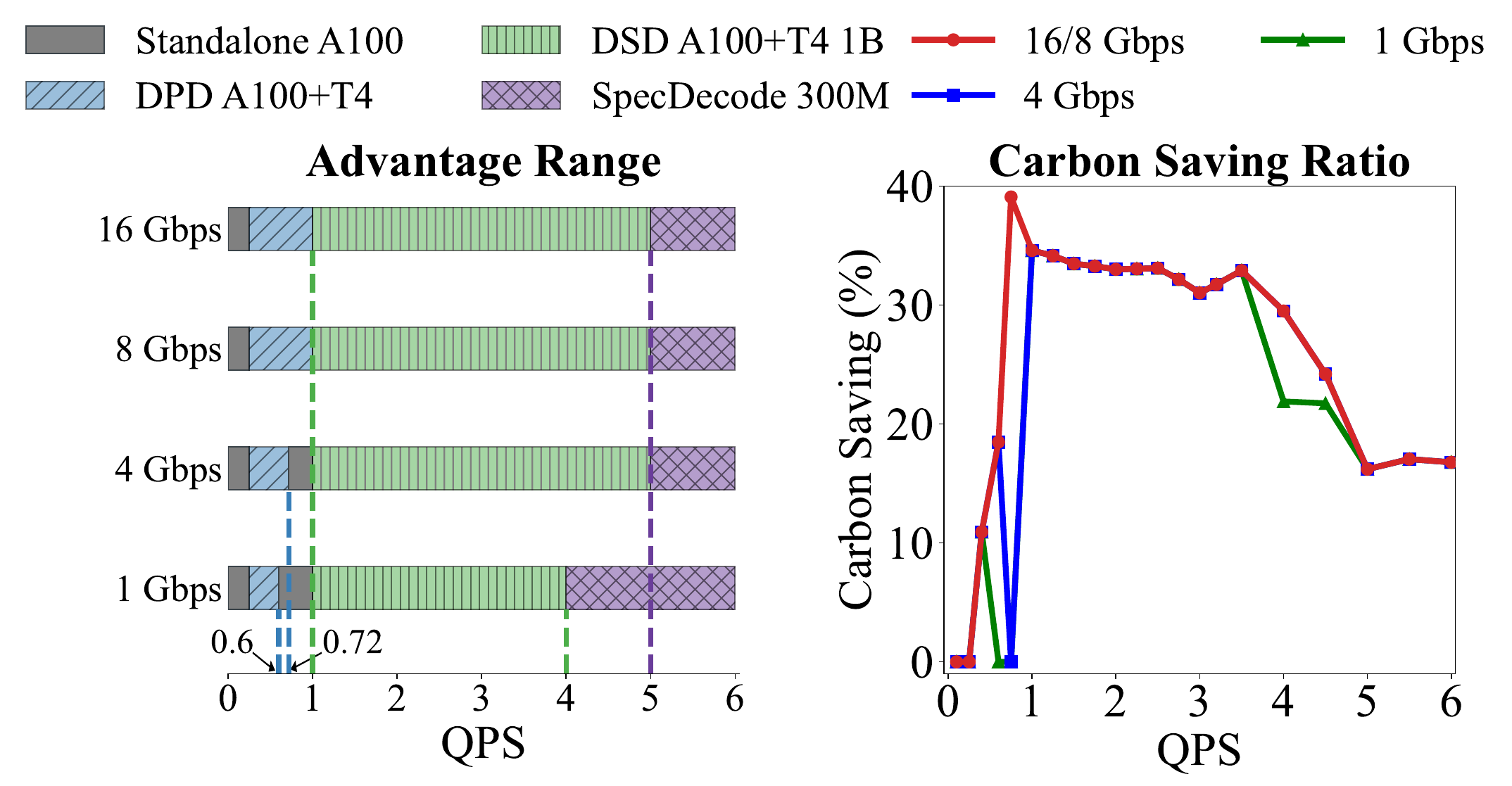}
    \caption{\textbf{Optimal configurations and the corresponding carbon savings of \SYSTEM{} across different QPS and network bandwidth.}}
    \label{fig:E-bandwidth}
\end{figure}

{
\setlength{\belowcaptionskip}{-3pt}
\begin{figure}[t]
    \centering
    \includegraphics[width=\linewidth]{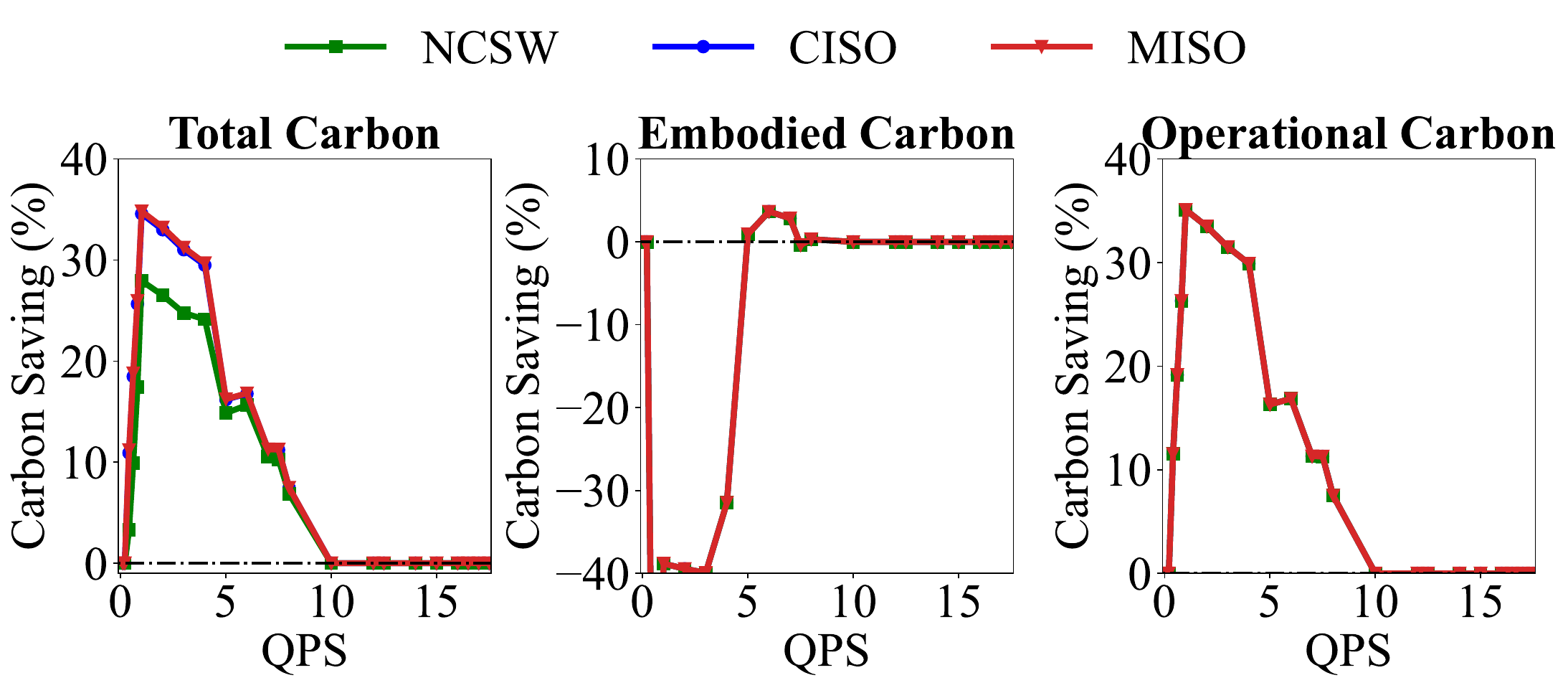}
    \caption{\textbf{Carbon saving of \SYSTEM{} in different geographical regions with varying carbon intensity.} NCSW, CISO, and MISO have carbon intensities of 17, 261, and 501 gCO$_{2}$/kWh, respectively.}
    \label{fig:E-CI}
\end{figure}
}
\subsection{Impact from Network Bandwidth}\label{subsec:res-bandwidth}

As discussed in~\Cref{subsec:bandwidth}, network bandwidth is crucial for efficient disaggregation and carbon reduction. This experiment investigates its impact on carbon savings by using ShareGPT dataset and controlling the request size to the median. 
The left side of \Cref{fig:E-bandwidth} summarizes the range of QPS under which each configuration achieves carbon savings across a range of network bandwidths, from 1 Gbps to 16 Gbps. The right side of \Cref{fig:E-bandwidth} shows the corresponding carbon savings. 
Across all bandwidths, \DistSD{} and SpecDecode consistently provide the highest carbon savings when QPS exceeds 1. In particular, SpecDecode performs best at a bandwidth of 1 Gbps. This is because these configurations do not require high bandwidth and, consequently, can operate effectively at high QPS rates. In contrast, \DistPD{} and Standalone, which rely on high bandwidth, are most effective at low QPS rates. Based on these findings, we conclude that when network bandwidth is limited, configurations involving speculative decoding, particularly disaggregated versions, can effectively use older GPUs.
% \todo{Add experimental setting description? Not mentioning Fig~\ref{fig:E-bandwidth} corresponds to ShareGPT median request size here.}

\subsection{Impact from Carbon Intensity}\label{subsec:res-ci}

To validate the consistency of \SYSTEM{}'s carbon savings across different geographical regions and to evaluate its performance under varying carbon intensity levels, we conduct an analysis using the ShareGPT P50 request size and three power grids: North Central Sweden (NCSW), California (CISO), and Midcontinent (MISO). These regions represent low, medium, and high carbon intensity levels, with carbon intensities of 17, 261, and 501 gCO$_{2}$/kWh, respectively. \Cref{fig:E-CI} shows the total carbon savings and a breakdown of these savings for \SYSTEM{} in each of these regions. Notably, \SYSTEM{} selects the same optimal configuration across all three regions, regardless of the significant differences in carbon intensity, leading to consistent embodied and operational carbon savings. As demonstrated by the analysis of~\Cref{eq:q2} in~\Cref{sec:analysis}, higher carbon intensity in CISO and MISO results in greater carbon savings compared to NCSW. However, even in regions with low carbon intensity, such as NCSW, \SYSTEM{} can still achieve carbon savings of up to 27.9\%. The close carbon savings observed between CISO and MISO suggest that once the carbon intensity reaches a sufficiently high level, further increases in carbon intensity have a marginal impact on the overall savings. In conclusion, \SYSTEM{} effectively translates improved energy efficiency into enhanced carbon efficiency across diverse geographical regions.

\subsection{Impact from GPU Lifetime}\label{subsec:res-lifetime}

We vary the expected lifetime of GPUs to show the impact on carbon emissions. 
Like before, we evaluate ShareGPT and control the input/output token length to its median. 
The ShareGPT result in \Cref{fig:E-carbon-main} shows two interesting points: one at QPS=1 where the disaggregation configuration \DistSD{} becomes the optimal carbon-saving configuration
and another at QPS=4 where it makes way for SpecDecode.
% stops to show benefits. 
\Cref{fig:E-lifetime} zooms into these two data points. 
The left-side figure shows the expected carbon savings change compared the default 7-year lifetime old GPUs to the old GPU's expected lifetime (5--10 years); the right-side shows the expected carbon savings change compared the default 7-year lifetime new GPUs to the new GPU's expected lifetime (2--7 years).
Overall, as the lifetime of old GPUs increases, the savings on carbon emissions increase.
In comparison, as the lifetime of new GPUs decreases, the savings on carbon emissions reduce. 
This trend is expected as when the old GPU's lifetime increases, its amortized embodied carbon per second decreases; when the new GPU's lifetime decreases, its amortized embodied carbon per second increases. 
The newly installed GPU (A100) in this paper has a shorter lifetime compared to older GPUs (T4 and V100) that have been in service for years.
This study validates the implications in \Cref{sec:analysis} that short-lifetime new GPUs can work cooperatively with old GPUs to reduce their embodied carbon emissions. 
% We can view the embodied carbon of old GPU in disaggregation systems as an additional overhead for adding more computing resources. Then we find that shorter lifetimes of new GPUs and longer lifetimes of old GPUs reduce the portion of this additional overhead in the total embodied carbon of the system.
% Therefore, we conclude that disaggregation yields the best carbon emission savings when using shorter-lifetime newer GPUs and longer-lifetime older GPUs. 

{
\setlength{\belowcaptionskip}{-3pt}
\begin{figure}[t]
    \centering
    \includegraphics[width=.8\linewidth]{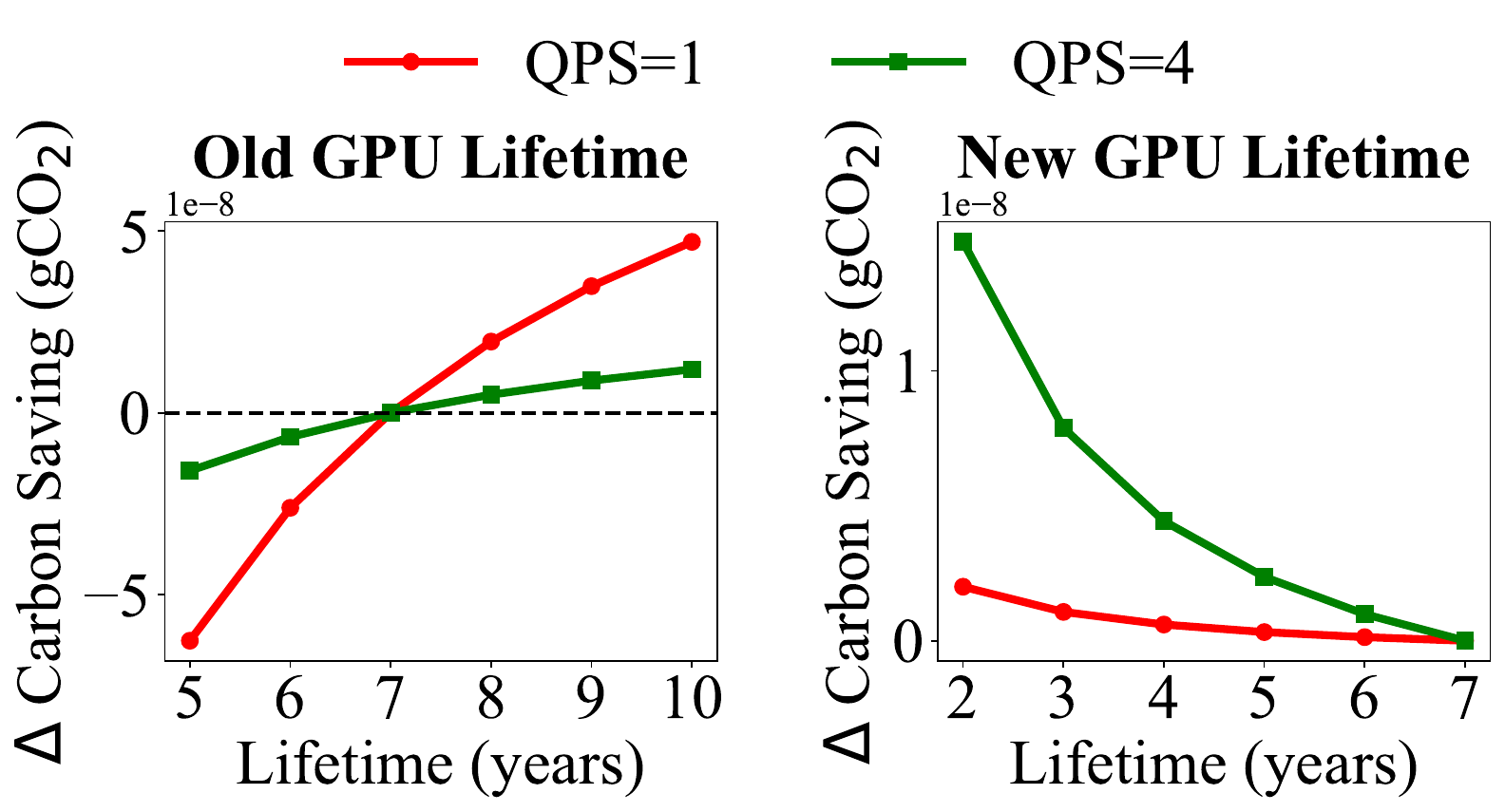}
    \caption{\textbf{Impact of GPU lifetime on carbon saving of DSD A100+T4 1B configuration}.}
    \label{fig:E-lifetime}
\end{figure}
}
\section{Conclusion} \label{sec:conclusion}

The wide adoption of LLMs has led to significant environmental impacts due to their high computational intensity and resource demands. This demand has driven the development of new generations of high-performing GPUs, exacerbating the issue of electronic waste and premature disposal of devices. \SYSTEM{} addresses the challenge of reducing the carbon emissions of LLM serving by reusing older, low-performing GPUs. By demonstrating the feasibility of reusing old GPUs and disaggregating workloads to balance carbon efficiency with performance goals, we hope to inspire future research into sustainable AI systems. Beyond LLM serving, the principles and methods introduced in \SYSTEM{} could extend to other compute-intensive AI applications, encouraging broader adoption of sustainable computing practices. 

% Ultimately, we aim for our work to spark innovation in environmentally conscious system design, fostering a more sustainable trajectory for AI's growth and its integration into society.

\bibliographystyle{plain}
\bibliography{reference}

\begin{thebibliography}{10}

\bibitem{ewaste2024}
The global e-waste monitor 2024.
\newblock {\em International Telecommunication Union (ITU) and United Nations
  Institute for Training and Research (UNITAR)}, 2024.

\bibitem{acun2023carbon}
Bilge Acun, Benjamin Lee, Fiodar Kazhamiaka, Kiwan Maeng, Udit Gupta, Manoj
  Chakkaravarthy, David Brooks, and Carole-Jean Wu.
\newblock Carbon explorer: {A} holistic framework for designing carbon aware
  datacenters.
\newblock In {\em Proceedings of the 28th ACM International Conference on
  Architectural Support for Programming Languages and Operating Systems
  (ASPLOS), Volume 2}, 2023.

\bibitem{databricks-llm-inference}
Megha Agarwal, Asfandyar Qureshi, Nikhil Sardana, Linden Li, Julian Quevedo,
  and Daya Khudia.
\newblock {LLM} inference performance engineering: Best practices.
\newblock
  \url{https://www.databricks.com/blog/llm-inference-performance-engineering-best-practices},
  2023.

\bibitem{agrawal2024taming}
Amey Agrawal, Nitin Kedia, Ashish Panwar, Jayashree Mohan, Nipun Kwatra,
  Bhargav Gulavani, Alexey Tumanov, and Ramachandran Ramjee.
\newblock Taming throughput-latency tradeoff in llm inference with
  sarathi-serve.
\newblock In {\em 18th USENIX Symposium on Operating Systems Design and
  Implementation (OSDI 24)}, 2024.

\bibitem{meta_llama}
Meta AI.
\newblock Llama: Open and efficient foundation language models.
\newblock \url{https://ai.facebook.com/blog/large-language-model-llama}, 2024.
\newblock Accessed: 2024-07-20.

\bibitem{Bai2023LongBenchAB}
Yushi Bai, Xin Lv, Jiajie Zhang, Hong Lyu, Jiankai Tang, Zhidian Huang,
  Zhengxiao Du, Xiao Liu, Aohan Zeng, Lei Hou, Yuxiao Dong, Jie Tang, and
  Juanzi Li.
\newblock Longbench: A bilingual, multitask benchmark for long context
  understanding.
\newblock {\em ArXiv}, abs/2308.14508, 2023.

\bibitem{ieee_spectrum_ewaste2024}
Katherine Bourzac.
\newblock Generative ai has a massive e-waste problem, 2024.

\bibitem{brakel2024model}
Felix Brakel, Uraz Odyurt, and Ana-Lucia Varbanescu.
\newblock Model parallelism on distributed infrastructure: A literature review
  from theory to llm case-studies, 2024.

\bibitem{Chen2021EvaluatingLL}
Mark Chen, Jerry Tworek, Heewoo Jun, Qiming Yuan, Henrique Pond{\'e}, Jared
  Kaplan, Harrison Edwards, Yura Burda, Nicholas Joseph, Greg Brockman, Alex
  Ray, Raul Puri, Gretchen Krueger, Michael Petrov, Heidy Khlaaf, Girish
  Sastry, Pamela Mishkin, Brooke Chan, Scott Gray, Nick Ryder, Mikhail Pavlov,
  Alethea Power, Lukasz Kaiser, Mohammad Bavarian, Clemens Winter, Philippe
  Tillet, Felipe~Petroski Such, David~W. Cummings, Matthias Plappert, Fotios
  Chantzis, Elizabeth Barnes, Ariel Herbert-Voss, William~H. Guss, Alex Nichol,
  Igor Babuschkin, Suchir Balaji, Shantanu Jain, Andrew Carr, Jan Leike, Joshua
  Achiam, Vedant Misra, Evan Morikawa, Alec Radford, Matthew~M. Knight, Miles
  Brundage, Mira Murati, Katie Mayer, Peter Welinder, Bob McGrew, Dario Amodei,
  Sam McCandlish, Ilya Sutskever, and Wojciech Zaremba.
\newblock Evaluating large language models trained on code.
\newblock {\em ArXiv}, abs/2107.03374, 2021.

\bibitem{chien2023reducing}
Andrew~A Chien, Liuzixuan Lin, Hai Nguyen, Varsha Rao, Tristan Sharma, and
  Rajini Wijayawardana.
\newblock Reducing the carbon impact of generative {AI} inference (today and in
  2035).
\newblock In {\em Proceedings of the 2nd Workshop on Sustainable Computer
  Systems (HotCarbon)}, 2023.

\bibitem{ai_e-waste_2024}
Casey Crownhart.
\newblock Ai will add to the e-waste problem. here's what we can do about it.
\newblock {\em MIT Technology Review}, 2024.

\bibitem{delimitrou2013paragon}
Christina Delimitrou and Christos Kozyrakis.
\newblock Paragon: Qos-aware scheduling for heterogeneous datacenters.
\newblock {\em ACM SIGPLAN Notices}, 48(4):77--88, 2013.

\bibitem{ding2019generative}
Yi~Ding, Nikita Mishra, and Henry Hoffmann.
\newblock Generative and multi-phase learning for computer systems
  optimization.
\newblock In {\em Proceedings of the 46th International Symposium on Computer
  Architecture}, 2019.

\bibitem{faiz2024llmcarbon}
Ahmad Faiz, Sotaro Kaneda, Ruhan Wang, Rita~Chukwunyere Osi, Prateek Sharma,
  Fan Chen, and Lei Jiang.
\newblock {LLMC}arbon: Modeling the end-to-end carbon footprint of large
  language models.
\newblock In {\em The Twelfth International Conference on Learning
  Representations}, 2024.

\bibitem{whyyourinternet2020}
Sarah Griffiths.
\newblock Why your internet habits are not as clean as you think.
\newblock
  \url{https://www.bbc.com/future/article/20200305-why-your-internet-habits-are-not-as-clean-as-you-think},
  2020.

\bibitem{griggs2024melange}
Tyler Griggs, Xiaoxuan Liu, Jiaxiang Yu, Doyoung Kim, Wei-Lin Chiang, Alvin
  Cheung, and Ion Stoica.
\newblock M\'elange: Cost efficient large language model serving by exploiting
  gpu heterogeneity, 2024.

\bibitem{gsteiger2024caribou}
Viktor~Urban Gsteiger, Pin~Hong Long, Yiran Sun, Parshan Javanrood, and
  Mohammad Shahrad.
\newblock Caribou: Fine-grained geospatial shifting of serverless applications
  for sustainability.
\newblock In {\em Proceedings of the ACM SIGOPS 30th Symposium on Operating
  Systems Principles}, 2024.

\bibitem{gupta2022act}
Udit Gupta, Mariam Elgamal, Gage Hills, Gu-Yeon Wei, Hsien-Hsin~S. Lee, David
  Brooks, and Carole-Jean Wu.
\newblock {ACT}: {Designing} sustainable computer systems with an architectural
  carbon modeling tool.
\newblock In {\em ISCA}, 2022.

\bibitem{gupta2021chasing}
Udit Gupta, Young~Geun Kim, Sylvia Lee, Jordan Tse, Hsien-Hsin~S Lee, Gu-Yeon
  Wei, David Brooks, and Carole-Jean Wu.
\newblock Chasing carbon: The elusive environmental footprint of computing.
\newblock In {\em 2021 IEEE International Symposium on High-Performance
  Computer Architecture (HPCA)}. IEEE, 2021.

\bibitem{hanafy2024carbonscaler}
Walid~A Hanafy, Qianlin Liang, Noman Bashir, David Irwin, and Prashant Shenoy.
\newblock Carbonscaler: leveraging cloud workload elasticity for optimizing
  carbon-efficiency.
\newblock {\em ACM SIGMETRICS Performance Evaluation Review}, 2024.

\bibitem{hanafy2024going}
Walid~A Hanafy, Qianlin Liang, Noman Bashir, Abel Souza, David Irwin, and
  Prashant Shenoy.
\newblock Going green for less green: Optimizing the cost of reducing cloud
  carbon emissions.
\newblock In {\em Proceedings of the 29th ACM International Conference on
  Architectural Support for Programming Languages and Operating Systems
  (ASPLOS)}, 2024.

\bibitem{hu2024TetriInfer}
Cunchen Hu, Heyang Huang, Liangliang Xu, Xusheng Chen, Jiang Xu, Shuang Chen,
  Hao Feng, Chenxi Wang, Sa~Wang, Yungang Bao, Ninghui Sun, and Yizhou Shan.
\newblock Inference without interference: Disaggregate {LLM} inference for
  mixed downstream workloads.
\newblock {\em arXiv preprint arXiv:2401.11181}, 2024.

\bibitem{kwon2023efficient}
Woosuk Kwon, Zhuohan Li, Siyuan Zhuang, Ying Sheng, Lianmin Zheng, Cody~Hao Yu,
  Joseph Gonzalez, Hao Zhang, and Ion Stoica.
\newblock Efficient memory management for large language model serving with
  pagedattention.
\newblock In {\em SOSP}, 2023.

\bibitem{leviathan2023fast}
Yaniv Leviathan, Matan Kalman, and Yossi Matias.
\newblock Fast inference from transformers via speculative decoding.
\newblock In {\em International Conference on Machine Learning}, 2023.

\bibitem{li2024uncertainty}
Amy Li, Sihang Liu, and Yi~Ding.
\newblock Uncertainty-aware decarbonization for datacenters.
\newblock In {\em Proceedings of the 3rd Workshop on Sustainable Computer
  Systems (HotCarbon)}, 2024.

\bibitem{li2024genai}
Baolin Li, Yankai Jiang, Vijay Gadepally, and Devesh Tiwari.
\newblock Toward sustainable {GenAI} using generation directives for
  carbon-friendly large language model inference, 2024.

\bibitem{li2024towards}
Yueying~Lisa Li, Omer Graif, and Udit Gupta.
\newblock Towards carbon-efficient llm life cycle.
\newblock In {\em Proceedings of the 3rd Workshop on Sustainable Computer
  Systems (HotCarbon)}, 2024.

\bibitem{liu2024your}
Jiawei Liu, Chunqiu~Steven Xia, Yuyao Wang, and Lingming Zhang.
\newblock Is your code generated by {ChatGPT} really correct? rigorous
  evaluation of large language models for code generation.
\newblock {\em Advances in Neural Information Processing Systems (NeurIPS)},
  2024.

\bibitem{maji2022carboncast}
Diptyaroop Maji, Prashant Shenoy, and Ramesh~K Sitaraman.
\newblock {CarbonCast: Multi}-day forecasting of grid carbon intensity.
\newblock In {\em Proceedings of the 9th ACM International Conference on
  Systems for Energy-Efficient Buildings, Cities, and Transportation
  (BuildSys)}, 2022.

\bibitem{miao2024specinfer}
Xupeng Miao, Gabriele Oliaro, Zhihao Zhang, Xinhao Cheng, Zeyu Wang, Zhengxin
  Zhang, Rae Ying~Yee Wong, Alan Zhu, Lijie Yang, Xiaoxiang Shi, et~al.
\newblock Specinfer: Accelerating large language model serving with tree-based
  speculative inference and verification.
\newblock In {\em Proceedings of the 29th ACM International Conference on
  Architectural Support for Programming Languages and Operating Systems, Volume
  3}, 2024.

\bibitem{nguyen2024towards}
Sophia Nguyen, Beihao Zhou, Yi~Ding, and Sihang Liu.
\newblock Towards sustainable large language model serving.
\newblock In {\em Proceedings of the 3rd Workshop on Sustainable Computer
  Systems (HotCarbon)}, 2024.

\bibitem{nvidia-llm-metrics}
Nvidia.
\newblock {NIM} for {LLM} benchmarking guide -- metrics.
\newblock
  \url{https://docs.nvidia.com/nim/benchmarking/llm/latest/metrics.html}, 2024.

\bibitem{ostrouchov2020gpulife}
George Ostrouchov, Don Maxwell, Rizwan~A. Ashraf, Christian Engelmann,
  Mallikarjun Shankar, and James~H. Rogers.
\newblock {GPU} lifetimes on titan supercomputer: {Survival} analysis and
  reliability.
\newblock In {\em Proceedings of the International Conference for High
  Performance Computing, Networking, Storage and Analysis (SC)}, 2020.

\bibitem{patel2024splitwise}
Pratyush Patel, Esha Choukse, Chaojie Zhang, Aashaka Shah, {\'I}{\~n}igo Goiri,
  Saeed Maleki, and Ricardo Bianchini.
\newblock Splitwise: Efficient generative {LLM} inference using phase
  splitting.
\newblock In {\em ISCA}, 2024.

\bibitem{patel2020clite}
Tirthak Patel and Devesh Tiwari.
\newblock Clite: Efficient and qos-aware co-location of multiple
  latency-critical jobs for warehouse scale computers.
\newblock In {\em 2020 IEEE International Symposium on High Performance
  Computer Architecture (HPCA)}, 2020.

\bibitem{protocol2011greenhouse}
Greenhouse~Gas Protocol.
\newblock Greenhouse gas protocol.
\newblock {\em Sector Toolsets for Iron and Steel-Guidance Document}, 2011.

\bibitem{radovanovic2022carbon}
Ana Radovanovi\'{c}, Ross Koningstein, Ian Schneider, Bokan Chen, Alexandre
  Duarte, Binz Roy, Diyue Xiao, Maya Haridasan, Patrick Hung, Nick Care, Saurav
  Talukdar, Eric Mullen, Kendal Smith, MariEllen Cottman, and Walfredo Cirne.
\newblock Carbon-aware computing for datacenters.
\newblock {\em IEEE Transactions on Power Systems}, 2022.

\bibitem{sharegpt}
{ShareGPT}.
\newblock Sharegpt - share and save your conversations with ai.
\newblock \url{https://sharegpt.com/}.

\bibitem{shen2024hugginggpt}
Yongliang Shen, Kaitao Song, Xu~Tan, Dongsheng Li, Weiming Lu, and Yueting
  Zhuang.
\newblock {HuggingGPT}: {Solving} {AI} tasks with {ChatGPT} and its friends in
  hugging face.
\newblock {\em Advances in Neural Information Processing Systems (NeurIPS)},
  2024.

\bibitem{shoeybi2019megatron}
Mohammad Shoeybi, Mostofa Patwary, Raul Puri, Patrick LeGresley, Jared Casper,
  and Bryan Catanzaro.
\newblock Megatron-lm: Training multi-billion parameter language models using
  model parallelism.
\newblock {\em arXiv preprint arXiv:1909.08053}, 2019.

\bibitem{souza2023ecovisor}
Abel Souza, Noman Bashir, Jorge Murillo, Walid Hanafy, Qianlin Liang, David
  Irwin, and Prashant Shenoy.
\newblock Ecovisor: {A} virtual energy system for carbon-efficient
  applications.
\newblock In {\em Proceedings of the 28th ACM International Conference on
  Architectural Support for Programming Languages and Operating Systems
  (ASPLOS), Volume 2}, 2023.

\bibitem{switzer2023junkyard}
Jennifer Switzer, Gabriel Marcano, Ryan Kastner, and Pat Pannuto.
\newblock Junkyard computing: Repurposing discarded smartphones to minimize
  carbon.
\newblock In {\em ASPLOS}, 2023.

\bibitem{vaswani2017attention}
Ashish Vaswani, Noam Shazeer, Niki Parmar, Jakob Uszkoreit, Llion Jones,
  Aidan~N Gomez, {\L}ukasz Kaiser, and Illia Polosukhin.
\newblock Attention is all you need.
\newblock {\em Advances in neural information processing systems}, 2017.

\bibitem{vu2023freshllms}
Tu~Vu, Mohit Iyyer, Xuezhi Wang, Noah Constant, Jerry Wei, Jason Wei, Chris
  Tar, Yun-Hsuan Sung, Denny Zhou, Quoc Le, and Thang Luong.
\newblock Freshllms: Refreshing large language models with search engine
  augmentation.
\newblock {\em Findings of the Association for Computational Linguistics ACL
  2024}, 2024.

\bibitem{wang2024designing}
Jaylen Wang, Daniel~S. Berger, Fiodar Kazhamiaka, Celine Irvene, Chaojie Zhang,
  Esha Choukse, Kali Frost, Rodrigo Fonseca, Brijesh Warrier, Chetan Bansal,
  Jonathan Stern, Ricardo Bianchini, and Akshitha Sriraman.
\newblock Designing cloud servers for lower carbon.
\newblock In {\em ISCA}, 2024.

\bibitem{wang2024waste}
Peng Wang, Ling-Yu Zhang, Asaf Tzachor, and Wei-Qiang Chen.
\newblock E-waste challenges of generative artificial intelligence.
\newblock {\em Nature Computational Science}, 2024.

\bibitem{chatgptcarbon2023}
Vinnie Wong.
\newblock {Gen AI}'s environmental ledger: {A} closer look at the carbon
  footprint of {ChatGPT}.
\newblock \url{https://piktochart.com/blog/carbon-footprint-of-chatgpt/}, 2023.

\bibitem{wu2022sustainable}
Carole-Jean Wu, Ramya Raghavendra, Udit Gupta, Bilge Acun, Newsha Ardalani,
  Kiwan Maeng, Gloria Chang, Fiona~Aga Behram, James Huang, Charles Bai,
  Michael~K. Gschwind, Anurag Gupta, Myle Ott, Anastasia Melnikov, Salvatore
  Candido, David Brooks, Geeta Chauhan, Benjamin Lee, Hsien-Hsin~S. Lee, Bugra
  Akyildiz, Maximilian Balandat, Joe Spisak, Ravi~Kumar Jain, Michael~G.
  Rabbat, and Kim~M. Hazelwood.
\newblock Sustainable ai: Environmental implications, challenges and
  opportunities.
\newblock {\em Proceedings of Machine Learning and Systems}, 2022.

\bibitem{yu2022orca}
Gyeong-In Yu, Joo~Seong Jeong, Geon-Woo Kim, Soojeong Kim, and Byung-Gon Chun.
\newblock Orca: A distributed serving system for transformer-based generative
  models.
\newblock In {\em 16th USENIX Symposium on Operating Systems Design and
  Implementation (OSDI)}, 2022.

\bibitem{zhong2024distserve}
Yinmin Zhong, Shengyu Liu, Junda Chen, Jianbo Hu, Yibo Zhu, Xuanzhe Liu, Xin
  Jin, and Hao Zhang.
\newblock Distserve: Disaggregating prefill and decoding for goodput-optimized
  large language model serving.
\newblock In {\em 18th USENIX Symposium on Operating Systems Design and
  Implementation (OSDI)}, 2024.

\end{thebibliography}

\end{document}